\documentclass[10pt,twocolumn,letterpaper]{article}
\usepackage[utf8]{inputenc}
\usepackage[margin=0.8in]{geometry}
\usepackage{setspace}
\usepackage{graphicx}
\usepackage{booktabs}
\usepackage{amsmath}%
\usepackage[labelfont=bf,skip=5pt,justification=justified, format=plain]{caption}
\usepackage[caption=false]{subfig}
\usepackage[style=chem-acs,articletitle=true,maxbibnames=30]{biblatex}
\usepackage [english]{babel}
\usepackage [autostyle, english = american]{csquotes}
\MakeOuterQuote{"}
\usepackage{lipsum}
\usepackage{url}

\begin{document}

\begin{singlespace}

\title{Mn(Pt$_{1-x}$Pd$_{x}$)$_5$P: Isovalent Tuning of Mn Sublattice Magnetic Order}
\author{Tyler J. Slade,$^{1,2||*}$ Ranuri S. Dissanayaka Mudiyanselage,$^{3||}$ Nao Furukawa,$^{1,2}$ Tanner R. Smith,$^{1,2}$ \\ Juan Schmidt,$^{1,2}$, Lin-Lin Wang,$^{1}$ Chang-Jong Kang,$^{4,5}$ Kaya Wei,$^{6}$ Zhixue Shu,$^{7}$ Tai Kong,$^{7}$ \\ Ryan Baumbach,$^{6}$ Gabriel Kotliar,$^{4}$ Sergey L. Bud’ko,$^{1,2}$ Weiwei Xie,$^{3,8}$ Paul C. Canfield$^{1,2*}$}
\date{}

\twocolumn[
\begin{@twocolumnfalse}

\maketitle

\begin{center} 
    
\textit{$^{1}$Ames National Laboratory, US DOE, Iowa State University, Ames, Iowa 50011, USA} \\  
\textit{$^{2}$Department of Physics and Astronomy, Iowa State University, Ames, Iowa 50011, USA} \\
\textit{$^{3}$Department of Chemistry and Chemical Biology, The State University of New Jersey Rutgers, Piscataway, NJ 08854, USA} \\
\textit{$^{4}$Department of Physics and Astronomy, Rutgers University, Piscataway, NJ, 08854, USA} \\
\textit{$^{5}$Department of Physics, Chungnam National University, Daejeon, 34134, South Korea} \\
\textit{$^{6}$National High Magnetic Field Laboratory, Tallahassee, FL, 32310, USA} \\
\textit{$^{7}$Department of Physics, University of Arizona, Tucson, AZ 85721, USA} \\
\textit{$^{8}$Department of Chemistry, Michigan State University, East Lansing, MI, 48824, USA} \\

\begin{abstract}

We report the growth and characterization of MnPd$_5$P, a rare-earth-free ferromagnet, with \textit{T}$_C$ $\approx$ 295 K and planar anisotropy, and conduct a substitutional study with its antiferromagnetic analogue MnPt$_5$P. We provide a solution route to grow large single crystals of MnPd$_5$P and the series Mn(Pt$_{1-x}$Pd$_x$)$_5$P by adding Mn into Pd-P and (Pt$_{1-x}$Pd$_{x}$)-P based melts. All compounds in the family adopt the layered anti-CeCoIn$_5$ type structure with the space group \textit{P}4/\textit{mmm}, and EDS and X-ray diffraction results indicate that MnPt$_5$P and MnPd$_5$P form a complete solid solution. Based on measurements of the temperature- and field-dependent magnetization and resistance, we construct a temperature-composition (\textit{T}–\textit{x}) phase diagram for Mn(Pt$_{1-x}$Pd$_x$)$_5$P and demonstrate that the initial antiferromagnetic order found in MnPt$_5$P is extraordinarily sensitive to Pd substitution. At low Pd fractions (\textit{x} $<$ 0.010), the single antiferromagnetic transition in pure MnPt$_5$P splits into a higher temperature ferromagnetic transition followed first, upon cooling, by a lower temperature ferromagnetic to antiferromagnetic transition and then by a re-entrant antiferromagnetic to ferromagnetic transition at even lower temperatures. The antiferromagnetic region makes up a bubble phase that persists up to \textit{x} $\approx$ 0.008–0.009 for \textit{T} $\approx$ 150 K, with all samples \textit{x} $<$ 0.008 recovering their initial ferromagnetic state upon further cooling to base temperature. Over the same low substitution range we find a non-monotonic change in the room temperature value of the unit cell volume, further suggesting that pure MnPt$_5$P is very close to an instability. Once \textit{x} $>$ 0.010, Mn(Pt$_{1-x}$Pd$_x$)$_5$P undergoes a only single transition into the ferromagnetic phase. The Curie temperature initially increases rapidly with \textit{x}, rising from \textit{T}$_C$ $\approx$ 197 K at \textit{x} = 0.013 to a maximum of \textit{T}$_C$ $\approx$ 312 K for \textit{x} $\approx$ 0.62, and then falling back to \textit{T}$_C$ $\approx$ 295 K for pure MnPd$_5$P (\textit{x} = 1.00). Given that Pt and Pd are isoelectronic, this work raises questions as to the origin of the extreme sensitivity of the magnetic ground state in MnPt$_5$P upon introducing Pd.

\end{abstract}


\end{center}

\vspace{7mm}

\end{@twocolumnfalse}
]

\section{Introduction}

Targeted design of tunable magnetic materials is active and key challenge for the materials chemistry and physics community. Achieving this goal necessitates understanding, at a microscopic level, what chemical and structural features underpin the magnetic properties of a given material. Whereas in magnetic semiconductors and insulators, theories based upon super-exchange interactions can often provide a satisfactory explanation of the magnetism,\autocite{dietl2010ten,goodenough1960direct} in metallic compounds, it is far more challenging from a theoretical perspective to understand and therefore to predict whether a structure containing transition metals will be paramagnetic, ferromagnetic, or antiferromagnetic.\autocite{samolyuk2008relation} This difficulty is particularly pronounced in the case of low-dimensional and/or itinerant magnetic metals, which show completely different electronic and magnetic properties from magnetic semiconductors.\autocite{fei2018two,zhang2017computational,shankhari2017unexpected,chen2013magnetic,fokwa2011rational} With these challenges in mind, detailed studies on isostructural or chemically similar intermetallic compounds with disparate magnetic properties may yield valuable insight into the physical parameters that ultimately determine the magnetism and hopefully provide guidelines for more targeted design of magnetic materials.

Based on our previous experimental work, magnetically active 3\textit{d} metals (most frequently Cr, Mn, Fe, Co and Ni) occupying voids in complex intermetallic frameworks can give rise to lower-dimensional structures of these magnetic metals. A prominent example is the \textit{M}(Pt, Pd)$_5$\textit{X} family, where \textit{M} = Mn, Fe and \textit{X} = P, As, Se.\autocite{gui2020chemical,gui2021spin,gui2020crystal,dissanayaka2022spin,SladeCanfieldZAAC} These compounds all crystalize in the layered CeCoIn$_5$-type tetragonal structure with the space group \textit{P}4/\textit{mmm} ($\#$123). Despite sharing the same crystal structure, the magnetic properties of the \textit{M}(Pt, Pd)$_5$\textit{X} materials are remarkably diverse. MnPt$_5$As is a \textit{T}$_C$ $\approx$ 280 K ferromagnet,\autocite{gui2020crystal} whereas the isovalent MnPt$_5$P orders antiferromagnetically below \textit{T}$_N$ $\approx$ 190 K (likely with a small ferromagnetic, \textit{q} = 0, component in addition to the predominantly antiferromagnetic order).\autocite{gui2020chemical,SladeCanfieldZAAC} FePt$_5$P is an itinerant antiferromagnet that undergoes three closely spaced transitions between $\approx$ 70-90 K.\autocite{gui2021spin,SladeCanfieldZAAC} Lastly, MnPd$_5$Se also shows antiferromagnetic order below \textit{T}$_N$ $\approx$ 80 K, and a spin reorientation is observed upon further cooling below 50 K.\autocite{dissanayaka2022spin} The range of properties exhibited by the \textit{M}(Pt, Pd)$_5$\textit{X} family suggests that the magnetism in these compounds is extremely sensitive to chemical composition, electron count, and steric considerations. The case of MnPt$_5$P is particularly interesting. As noted above, MnPt$_5$P orders antiferromagnetically, whereas MnPt$_5$As is a $\approx$ 280 K ferromagnet. Very recently, exploratory solution growth of single crystal studies found that isovalent MnPd$_5$P also manifests ferromagnetic order near room temperature,\autocite{SladeCanfieldZAAC} suggesting that both lattice expansion (towards MnPt$_5$As) and contraction (towards MnPd$_5$P) push the antiferromagnetic MnPt$_5$P toward a ferromagnetic ground state. 

In this work, we present a substitutional study between MnPd$_5$P and MnPt$_5$P to better characterize the magnetism of the end members and to understand how the initially antiferromagnetic state of MnPt$_5$P evolves towards ferromagnetism in MnPd$_5$P. We successfully grow single and polycrystalline samples of MnPd$_5$P and Mn(Pt$_{1-x}$Pd$_x$)$_5$P using both solution growth and solid-state reaction techniques. Phase and structural analysis with X-ray diffraction and energy dispersive spectroscopy indicate that MnPt$_5$P and MnPd$_5$P form a full solid solution of Mn(Pt$_{1-x}$Pd$_x$)$_5$P. Magnetization measurements show that the essentially antiferromagnetic state in pure MnPt$_5$P is extraordinarily sensitive to Pd substitution. At Pd concentrations as low as \textit{x} $<$ 0.01, the single antiferromagnetic transition found for pure MnPt$_5$P splits into a higher temperature ferromagnetic transition followed first, upon cooling, by a lower temperature ferromagnetic to antiferromagnetic transition and then by a re-entry into the ferromagnetic state at lower temperatures. The antiferromagnetic region makes up a bubble phase that persists up to \textit{x} $\approx$ 0.008–0.009 for \textit{T} $\approx$ 150 K, with all samples \textit{x} $<$ 0.008 recovering their initial ferromagnetic state upon further cooling to base temperature.  Over the same low substitution range we find a possible non-monotonic change in the room temperature value of the \textit{a}-lattice parameter and unit cell volume, further suggesting that pure MnPt$_5$P is very close to an instability. When \textit{x} $>$ 0.010, Mn(Pt$_{1-x}$Pd$_x$)$_5$P undergoes only a single, ferromagnetic, transition, where the Curie temperature initially increases with \textit{x} and is maximized at $\approx$ 312 K for \textit{x} = 0.62 Pd before decreasing to $\approx$ 295 K in pure MnPd$_5$P. Considering that for \textit{x} $>$ 0.01, the rather gradual and non-monotonic \textit{T}$_C$ variation does not track well with lattice parameters, or suggest particular sensitivity to Pd content, the fantastic sensitivity of the antiferromagnetic phase of Mn(Pt$_{1-x}$Pd$_x$)$_5$P for \textit{x} $<$ 0.01 is remarkable and suggests there is a qualitative change in the material that may well be associated with a change in the Fermi-surface topology, such as a Lifshitz transition. Qualitatively in-line with this proposal, electronic band structure calculations for MnPt$_5$P indeed show several pockets near the Fermi level which are significantly altered in MnPd$_5$P. Direct experiments to probe the electronic structure/density of states at the Fermi level are needed to further understand the extraordinary sensitivity of MnPt$_5$P to Pd alloying.

\section{Experimental Details}

\subsection{Crystal Growth} We prepared and characterized both single crystalline and polycrystalline samples of MnPd$_5$P and Mn(Pt$_{1-x}$Pd$_x$)$_5$P. The Mn(Pt$_{1-x}$Pd$_x$)$_5$P single crystals were grown from (Pt$_{1-x}$Pd$_x$)-P based solutions as follows.\autocite{SladeCanfieldZAAC} Elemental Mn pieces (Puratronic, 99.98$\%$), Pt powder (Engelhard, 99+ $\%$ purity), Pd powder (Engelhard, 99+ $\%$ purity), and red P pieces (Alpha-Aesar, 99.99$\%$) were weighed according to nominal compositions of Mn$_9$Pt$_{71-y}$Pd$_y$P$_{20}$ (the actual compositions are given in Table \ref{EDS_SC} in the Appendix) and placed into the bottom of an alumina Canfield crucible set (CCS).\autocite{canfield2016use,lspceramics_CCS} The packed CCS were flame sealed into evacuated fused silica ampules that were backfilled with $\approx$ 1/6 atm Ar gas. Using a box furnace, the ampules were slowly warmed to 250°C over 6 h and then to 1180°C over an additional 8 h. After dwelling at 1180°C for 6 h, the furnace was gradually cooled to 800°C (for samples with nominally less than 50$\%$ Pd) or to 830°C (for those with over 50$\%$ Pd) over $\approx$ 100 h. Upon reaching the desired temperature, the excess liquid phase was decanted by inverting the ampules into a specially designed centrifuge with metal rotor and cups.\autocite{canfield2019new} After cooling to room temperature, the ampules and CCS were opened to reveal clusters of metallic plate-like crystals with typical dimensions of $\approx$ 3 mm, and a representative picture of several crystals is shown in the inset to Figure \ref{PXRD}c.

The polycrystalline samples were prepared from a solid state reaction by sintering pellets with nominal compositions of Mn(Pt$_{1-x}$Pd$_x$)$_5$P (x = 0.2, 0.4, 0.5, 0.6, 0.8, and 1). Mn powder (Mangan, 99+$\%$), Pt powder (BTC, 22 mesh, 99.99$\%$), Pd powder (BTC, 200 mesh, 99.95$\%$) and red P powder (BTC, 100 mesh, 99$\%$) were mixed and ground in Mn: Pt/Pd: P = 1:5:1 atomic ratio. The mixture was pressed into a pellet, and the pellet was placed into an alumina crucible and sealed in an evacuated silica tube. The sample tube was then heated to 1050°C at a rate of 40 °C per hour. After annealing for 2 days at 1050°C, the samples were slowly cooled down to room temperature at the speed of 10°C per hour. Both the single- and polycrystalline Mn(Pt$_{1-x}$Pd$_x$)$_5$P samples were stable in moist air.

\subsection{Phase and Structure Determination} 
\begin{figure*}[!t]
    \centering
    \includegraphics[width=\linewidth]{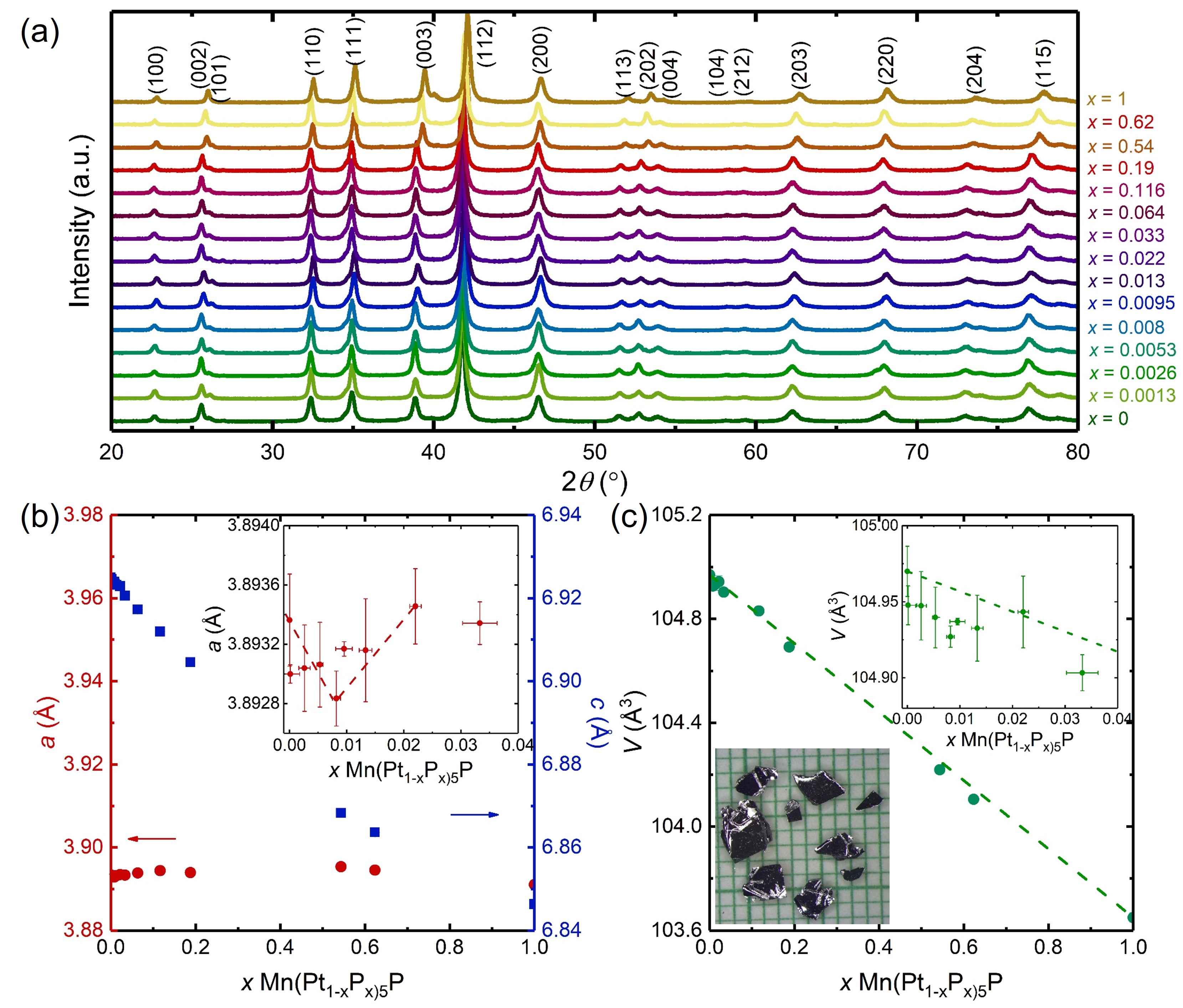}
    \caption[]{(a) Powder  X-ray diffraction patterns collected from the solution-grown Mn(Pt$_{1-x}$Pd$_x$)$_5$P crystals. The Pd fractions were determined from EDS. (b) Refined lattice parameters. The inset shows a close up view of the \textit{a}-lattice parameter for \textit{x} $<$ 0.04 samples, and the red dashed line is a guide to the eye showing possible non-monotonic behavior. (b) Refined unit cell volume. The upper inset is a zoomed in view of \textit{V} for \textit{x} $<$ 0.04 samples and the lower inset show representative crystals (\textit{x} = 0.064) on a mm grid. In (c) the dashed green line is a line connecting \textit{x} = 0 and \textit{x} = 1, showing overall close agreement of Vegard's law with possible deviation at very low \textit{x}. The horizontal error bars in (b) and (c) are the standard deviations from the EDS measurements, and the vertical error bars are discussed in the powder diffraction section of the experimental section.}
    \label{PXRD}
\end{figure*}

Powder X-ray diffraction patterns were obtained using a Rigaku Miniflex-II instrument operating with Cu-\textit{K}$\alpha$ radiation with $\lambda$ = 1.5406 \AA\ (\textit{K}$\alpha$1) and 1.5443 \AA\ (\textit{K}$\alpha$2) at 30 kV and 15 mA. The samples were prepared by grinding a representative number of crystals (5-10) to a fine powder. To determine the lattice parameters, the powder patterns were refined using the Rietveld method with GSAS-II software.\autocite{toby2013gsas} To obtain better estimate of the uncertainty in the lattice parameters for samples very dilute in Pd, we collected and refined three separate patterns for samples with \textit{x} $\leq$ 0.022, and used the standard deviations of the refined values of \textit{a}, \textit{c}, and \textit{V} as error bars (see Figure \ref{PXRD}b and \ref{PXRD}c). For samples with \textit{x} $>$ 0.022, the fitting errors from the Rietveld refinements were used as the error bars.


Single crystal X-ray diffraction (SCXRD) experiments were conducted in a D8 Quest Eco diffractometer with Mo-\textit{K}$\alpha$ radiation ($\lambda$ = 0.71073 \AA) equipped with Photon II detector. Empirically, we found that the solution grown single crystals of Mn(Pt$_{1-x}$Pd$_x$)$_5$P were not favorable for SCXRD, whereas very small single crystals picked from the sintered pellets (the solid-state reactions) were more suitable and were used for the SCXRD. The samples were mounted on a Kapton loop and measured with an exposure time of 10 s per frame scanning 2$\theta$ width of 0.5°. Structure refinement was performed in SHELXTL package using direct methods and full matrix least-squares on \textit{F}$^2$ model.\autocite{sheldrick2015crystal,muller2006crystal} Anisotropic thermal parameters for all atoms were refined in SHELXTL. The VESTA software was used to plot the crystal structures.\autocite{momma2008vesta}

\subsection{Scanning Electron Microscopy (SEM) and Elemental Analysis} The Pd concentrations (\textit{x}) in the Mn(Pt$_{1-x}$Pd$_x$)$_5$P single crystals were determined by energy dispersive x-ray spectroscopy (EDS) quantitative chemical analysis using an EDS detector (Thermo NORAN Microanalysis System, model C10001) attached to a JEOL scanning-electron microscope (SEM). The compositions of each crystal were measured at several (3–6) different positions on the crystal’s face (perpendicular to the \textit{c}-axis), revealing good homogeneity in each crystal. An acceleration voltage of 16 kV, working distance of 10 mm, and take off angle of 35° were used for measuring all standards and samples. A pure MnPt$_5$P single crystal (\textit{x} = 0.000) was used as a standard for Mn, Pt, and P quantification, and a pure MnPd$_5$P single crystal (\textit{x} = 1.000) was used as a standard for Pd. The spectra were fitted using NIST-DTSA II Microscopium software.\autocite{newbury2014rigorous} The average compositions and error bars were obtained from these data, accounting for both inhomogeneity and goodness of fit of each spectra. Chemical compositions of the polycrystalline Mn(Pt$_{1-x}$Pd$_x$)$_5$P samples were analyzed using a high vacuum Zeiss Sigma Field Emission SEM (FESEM) with Oxford INCA PentaFETx3 Energy-Dispersive Spectroscopy (EDS) system. Spectra were collected for 100 s from multiple areas of the crystals mounted on a carbon tape with an accelerating voltage of 20 keV.

\subsection{Physical Property Measurements} For the single crystals, magnetization measurements were performed in a Quantum Design Magnetic Property Measurement System (MPMS-classic) SQUID magnetometer operating in the DC measurement mode. The magnetic measurements were conducted with the field oriented parallel and perpendicular to the \textit{c}-axis, where \textit{c} is axial direction relative to the plate-like crystals. For measurements with \textit{H} $\perp$ \textit{c}, the samples were held in place between two plastic straws, and for \textit{H} $\parallel$ \textit{c}, the samples were sandwiched from above and below between two plastic discs. A small pinhole was poked in the space between the discs to allow for evacuation. In the latter case, a blank background using the bare discs was first measured and the values subtracted.

The temperature dependent resistance of the Mn(Pt$_{1-x}$Pd$_x$)$_5$P single crystals were measured on a Quantum Design Physical Property Measurement System (PPMS) operating in the AC transport mode with an applied current of 3 mA and frequency of 17 Hz. The samples were prepared by cutting the crystals into rectangular bars, and the contacts were made by spot welding 25 $\mu$m thick annealed Pt wire onto the samples in standard four point geometry. After spot welding, a small amount of silver epoxy was painted onto the contacts to ensure good mechanical strength, and typical contact resistances were $\approx$ 1 $\Omega$.

\subsection{Computational Details}

Electronic band structure and density of states for MnPt$_5$P and MnPd$_5$P were calculated in density functional theory (DFT)\autocite{hohenberg1964inhomogeneous,PhysRev.140.A1133} using PBE\autocite{PhysRevLett.100.136406} as the exchange-correlation functional with spin-orbit coupling (SOC) included. All DFT calculations were performed in the Vienna Ab initio Simulation Package (VASP)\autocite{PhysRevB.54.11169,KRESSE199615} with a plane-wave basis set and projector augmented wave method.\autocite{PhysRevB.50.17953} The kinetic energy cutoff was 270 eV. We used a $\Gamma$-centered Monkhorst-Pack\autocite{PhysRevB.13.5188} (10$\times$10$\times$6) \textit{k}-point mesh with a Gaussian smearing of 0.05 eV for the primitive tetragonal unit cell. 

 
\section{Results and Discussion}

\begin{figure}[!t]
    \centering
    \includegraphics[width=\linewidth]{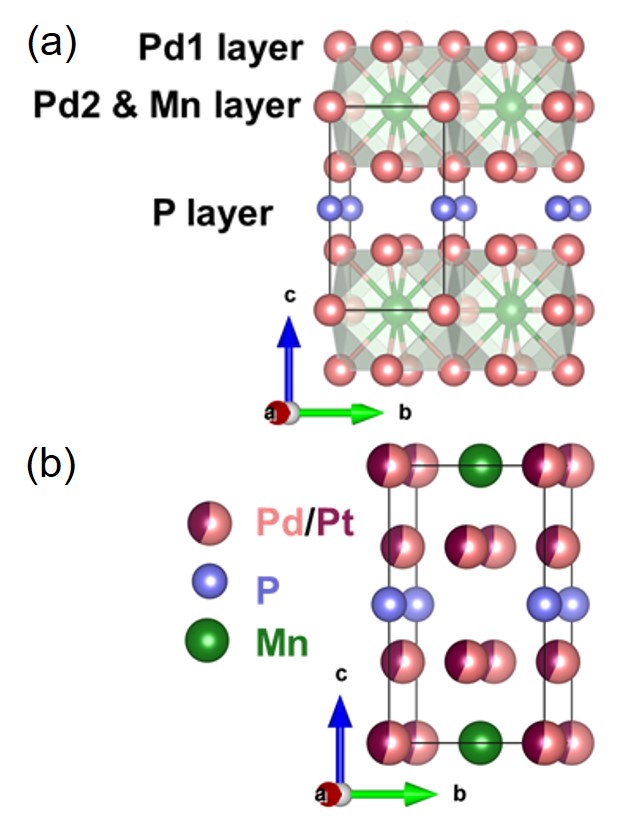}
    \caption[]{(a) The crystal structure of MnPd$_5$P showing Mn@Pd$_{12}$ face sharing polyhedral and P layers (b) Unit cell of Mn(Pt$_{1-x}$Pd$_x$)$_5$P indicating the mixture of Pd and Pt.}
    \label{structure}
\end{figure}

\begin{table}[htbp]
  \centering
  \caption{Single crystal structure refinement details for MnPd$_5$P at 300(2) K.}
    \begin{tabular}{ll}
    \toprule
    Refined Formula & MnPd$_5$P \\
    F.W. (g/mol) & \multicolumn{1}{r}{617.91} \\
    Space group; \textit{Z} & \textit{P}4/\textit{mmm}; 1 \\
    \textit{a} (\AA) & 3.899 (2) \\
    \textit{c} (\AA) & 6.867 (4) \\
    \textit{V} (\AA$^3$) & 104.42 (9) \\
    $\theta$ range (º) & 2.966-34.770 \\
    No. reflections; \textit{R}$_{int}$ & 578; 0.0609 \\
    No. independent reflections & \multicolumn{1}{r}{170} \\
    No. parameters & \multicolumn{1}{r}{12} \\
    \textit{R}$_1$: w\textit{R}$_2$ (\textit{I} $>$ 2\textit{d}(\textit{I})) & 0.0509; 0.1204 \\
    Goodness of fit & \multicolumn{1}{r}{1.282} \\
    Diffraction peak and hole (e$^-$/\AA$^3$) & 2.656; -1.863 \\
    \bottomrule
    \end{tabular}%
  \label{SCrefinement}%
\end{table}%

\begin{table}[t]
  \small
  \centering
  \caption{Atomic coordinates and equivalent isotropic displacement parameters of MnPd$_5$P at 300(2) K. (\textit{U}$_{eq}$ is defined as one-third of the trace of the orthogonalized \textit{U}$_{ij}$ tensor (\AA$^2$))}
    \resizebox{\linewidth}{!}{\begin{tabular}{llrrrrl}
    \toprule
    Atoms & Wycoff & \multicolumn{1}{l}{Occ.} & \multicolumn{1}{l}{\textit{x}} & \multicolumn{1}{l}{\textit{y}} & \multicolumn{1}{l}{\textit{z}} & \textit{U} (eq) \\
    \midrule
    Pd1 & 4\textit{i}  & 1   & 0   & \multicolumn{1}{l}{1/2} & 0.2948(1) & 0.015(1) \\
    Pd2 & 1\textit{a}  & 1   & 0   & 0   & 0 & 0.012(2) \\
    Mn3   & 1\textit{c}  & 1   & \multicolumn{1}{l}{1/2} & \multicolumn{1}{l}{1/2} & 0 & 0.022(2) \\
    P4   & 1\textit{b}  & 1   & 0   & 0   & \multicolumn{1}{r}{1/2}   & 0.016(2) \\
    \bottomrule
    \end{tabular}}%
  \label{AtomsU}%
\end{table}%

\begin{figure*}[!t]
    \centering
    \includegraphics[width=\linewidth]{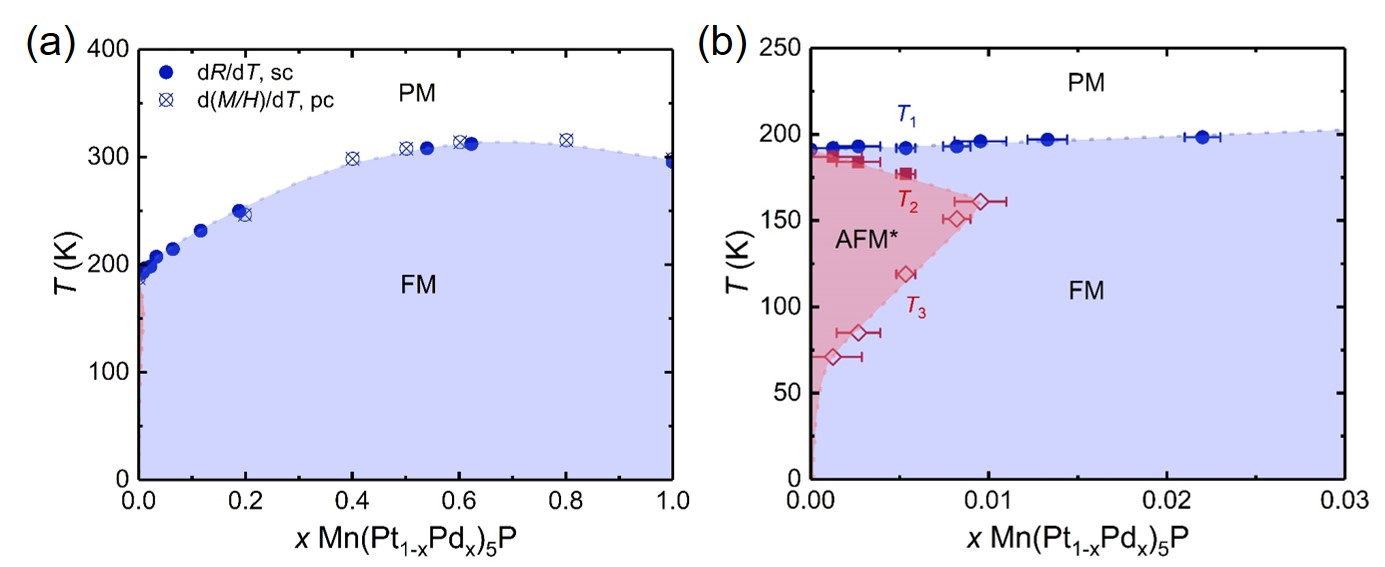}
    \caption[]{(a) Full temperature–composition phase diagram for Mn(Pt$_{5-x}$Pd$_x$)$_5$P. (b) The low–\textit{x} region of the phase diagram for \textit{x} $<$ 0.03. The * next to AFM denotes the small ferromagnetic (\textit{q} = 0) component to the primarily antiferromagnetic order in the low \textit{x} samples and the dashed lines are guides to the eye. In (a), the closed points represented dated obtained from the single crystals (sc) and the crossed-open points data from the polycrystalline (pc) samples.}
    \label{T-x}
\end{figure*}

\subsection{Phase, Composition, and Structural Analysis} We first analyzed our solution grown single crystals and sintered pellets with powder X-ray diffraction (PXRD) and energy dispersive spectroscopy (EDS) to determine the phase and assess the degree of Pd incorporation into the Mn(Pt$_{1-x}$Pd$_x$)$_5$P alloys.  Figure \ref{PXRD}a shows the powder patterns collected for the ground, solution grown single crystals, and the EDS data is given in Table \ref{EDS_SC} in the Appendix. For all samples, the experimental PXRD patterns are in excellent agreement with the anticipated reflections for the \textit{P}4/\textit{mmm} structure of MnPt$_5$P and MnPd$_5$P. The EDS analysis likewise suggests a monotonic increase in the Pd incorporation into the 1-5-1 matrix as Pt is exchanged for Pd in the starting melts (see Table \ref{EDS_SC}).

The lattice parameters determined from Rietveld refinements of the powder patterns are shown in Figure \ref{PXRD}b. The \textit{a}-lattice parameter has a very shallow maximum at \textit{x} = 0.054, but overall there is little change in \textit{a} over the full compositional range. This is contrasted by the \textit{c}-lattice parameter which decreases monotonically (nearly linearly) as the Pd fraction increases. Because \textit{a} is nearly invariant with Pd doping, the unit cell volume \textit{V}, shown in Figure \ref{PXRD}c, essentially mirrors the Pd dependence of \textit{c}, decreasing linearly as the Pd content is raised. The dashed green line in Figure \ref{PXRD}c shows a linear fit between the volume of \textit{x} = 0 and \textit{x} = 1, and the experimental values closely follow the projected line, indicating that \textit{V} follows Vegard’s law for a solid solution between MnPt$_5$P and MnPd$_5$P. At very low Pd fraction (\textit{x} $<$ 0.22) the \textit{a} lattice parameter and unit cell volume arguably each have a V-shape \textit{x}-dependency, initially decreasing slightly before increasing again (see insets to Figure \ref{PXRD}b and \ref{PXRD}c). This non-monotonic is more evident in the volume, where a clear deviation from the Vegard's law is observed for \textit{x} $<$ 0.22. We will return to this anomalous \textit{x}-dependency at low substitution in the discussion of the magnetic properties.

To provide more detailed structural analysis of MnPd$_5$P and the Mn(Pt$_{1-x}$Pd$_x$)$_5$P alloys, we conducted single crystal X-ray diffraction (SCXRD). The resulting crystallographic data, including atomic coordinates, site occupancies and equivalent isotropic thermal displacement parameters of MnPd$_5$P, are reported in Table \ref{SCrefinement} and Table \ref{AtomsU} whereas crystallographic information on the Mn(Pt$_{1-x}$Pd$_x$)$_5$P alloys are given in the Appendix in Tables \ref{Mn(Pt1-xPdx)5P_crystallography} and \ref{Mn(Pt1-xPdx)P_refineatoms}. The results show that MnPd$_5$P and the Mn(Pt$_{1-x}$Pd$_x$)$_5$P compounds crystalize in a tetragonal unit cell with the space group of \textit{P}4/\textit{mmm}, like the previously reported MnPt$_5$P and MnPt$_5$As. The crystal structure is illustrated in Figures \ref{structure}a and \ref{structure}b and consists of layered motifs, with alternating layers of Mn@Pd$_{12}$ face sharing polyhedra that span the \textit{ab}-plane and which are separated by P layers along the \textit{c}-axis. Consistent with the powder diffraction data, the single crystal XRD confirms that the Mn(Pt$_{1-x}$Pd$_x$)$_5$P alloys maintain the parent lattice structure with the Pt and Pd atoms having mixed occupancy on the two atomic sites 1\textit{a} and 4\textit{i} as indicated in Figure \ref{structure}b. Details on the Pt/Pd distributions on 1\textit{a} and 4\textit{i} sites for each phase are given in Table \ref{Mn(Pt1-xPdx)P_refineatoms} in the Appendix. The SCXRD data may hint that the Pd atoms have a slight preference for occupying the 4\textit{i} site over the 1\textit{a} site; however, given the uncertainties in our refinements, this cannot be supported with confidence and our data indicates the Pt/Pd mixing is essentially a solid solution.


\begin{figure}[!t]
    \centering
    \includegraphics[width=\linewidth]{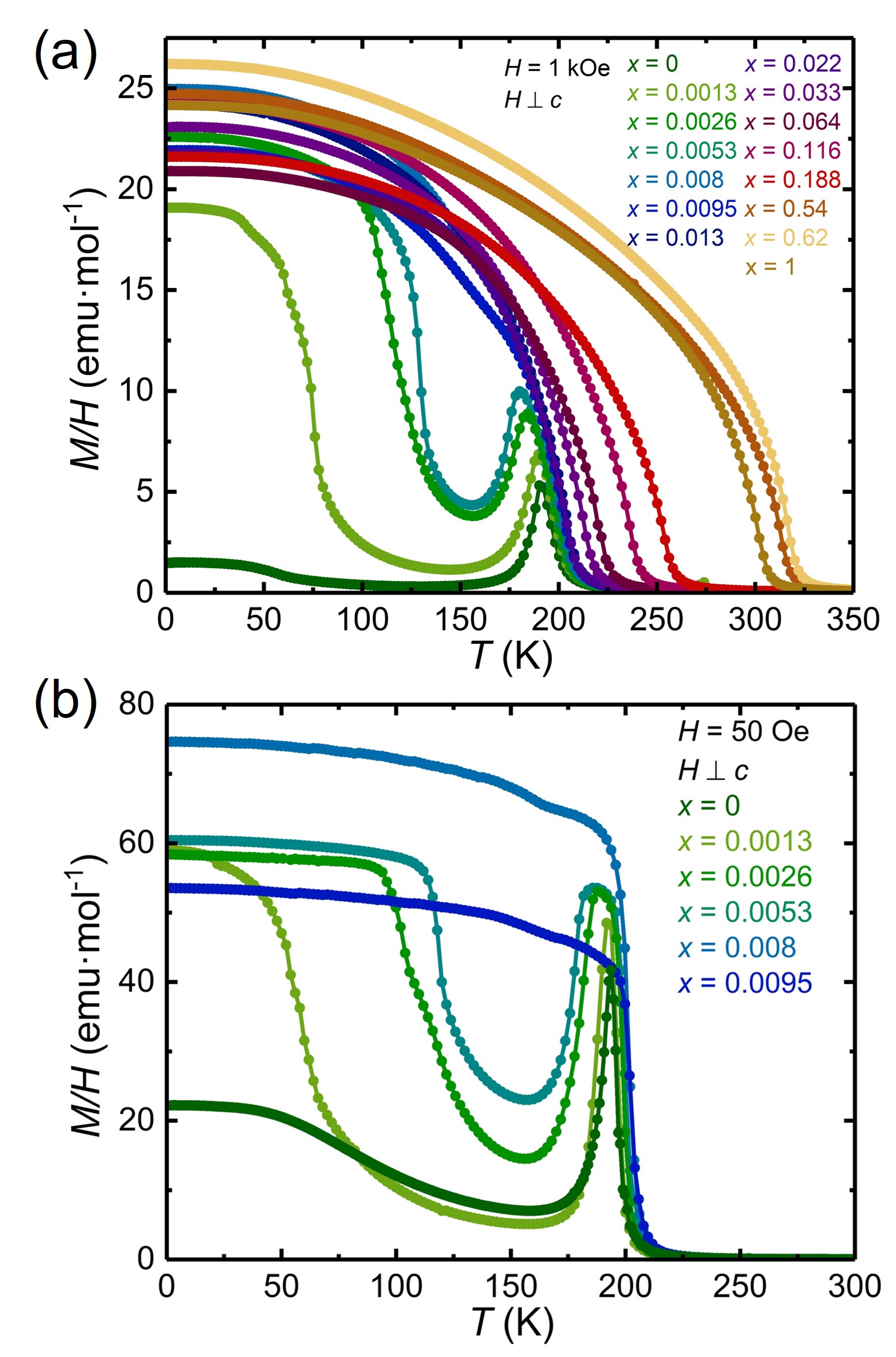}
    \caption[]{(a) Temperature dependence of \textit{M/H} for all Mn(Pt$_{1-x}$Pd$_x$)$_5$P single crystals measured at \textit{H} = 1 kOe. (b) Temperature dependence of \textit{M/H} for low \textit{x} Mn(Pt$_{1-x}$Pd$_x$)$_5$P with \textit{x} $<$ 1 measured at \textit{H} = 50 Oe.}
    \label{MT}
\end{figure}

\begin{figure}[!t]
    \centering
    \includegraphics[width=\linewidth]{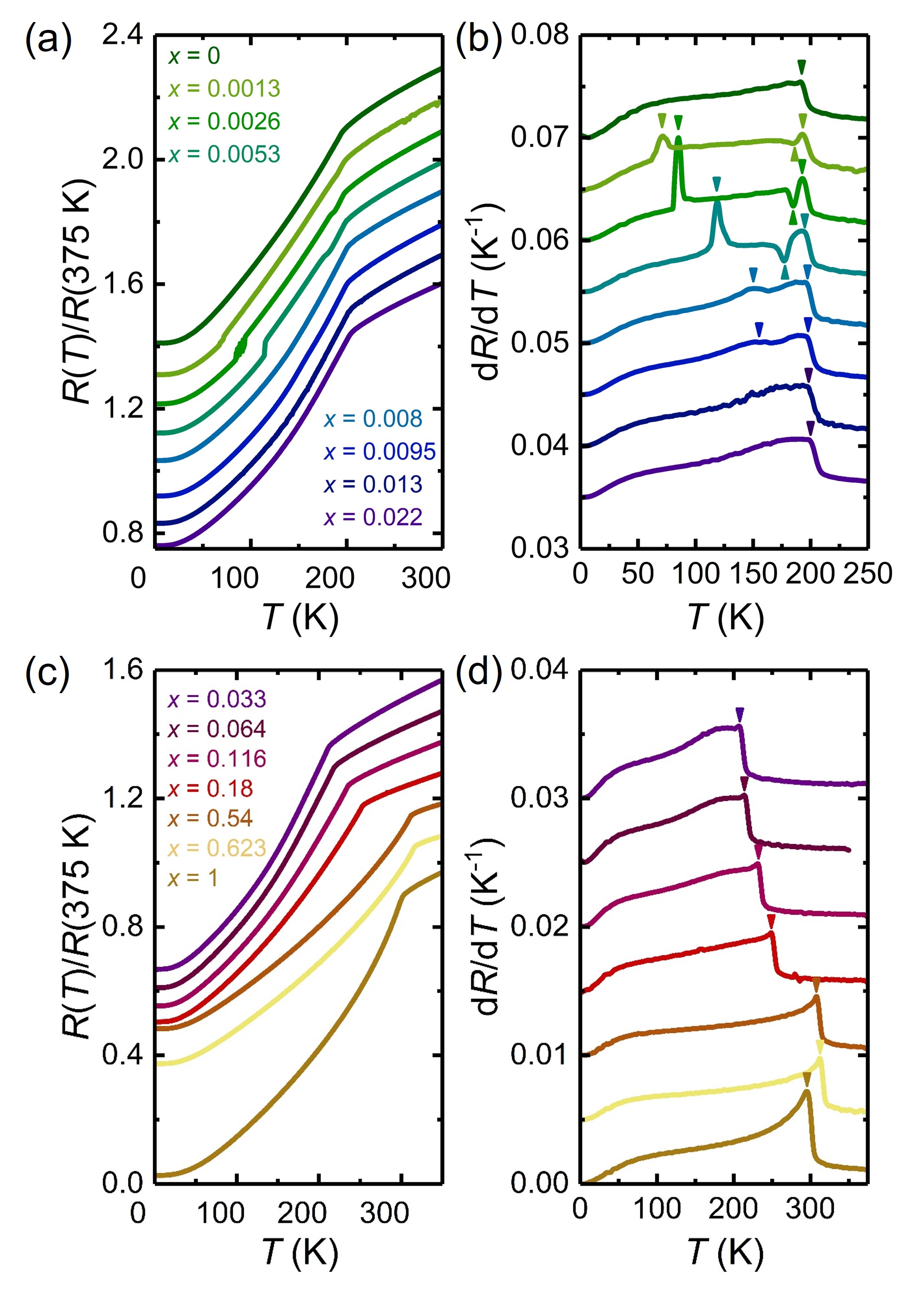}
    \caption[]{(a) Temperature dependent resistance data for 0 $\geq$ \textit{x} $\geq$ 0.022 Mn(Pt$_{1-x}$Pd$_x$)$_5$P single crystals with the data normalized to \textit{R}(375 K). For clarity, the \textit{R}(\textit{T}) curves are each offset by 0.1. (b) Derivatives of the datasets in (a). The peaks were used to determine transition temperatures and are marked with arrows.\autocite{PhysRevLett.20.665} (c) and (d) are the respective \textit{R}(\textit{T})/\textit{R}(375 K) and d\textit{R}/d\textit{T} data for 0.033 $\geq$ x $\geq$ 1 }
    \label{RT}
\end{figure}

\begin{figure*}[!t]
    \centering
    \includegraphics[width=\linewidth]{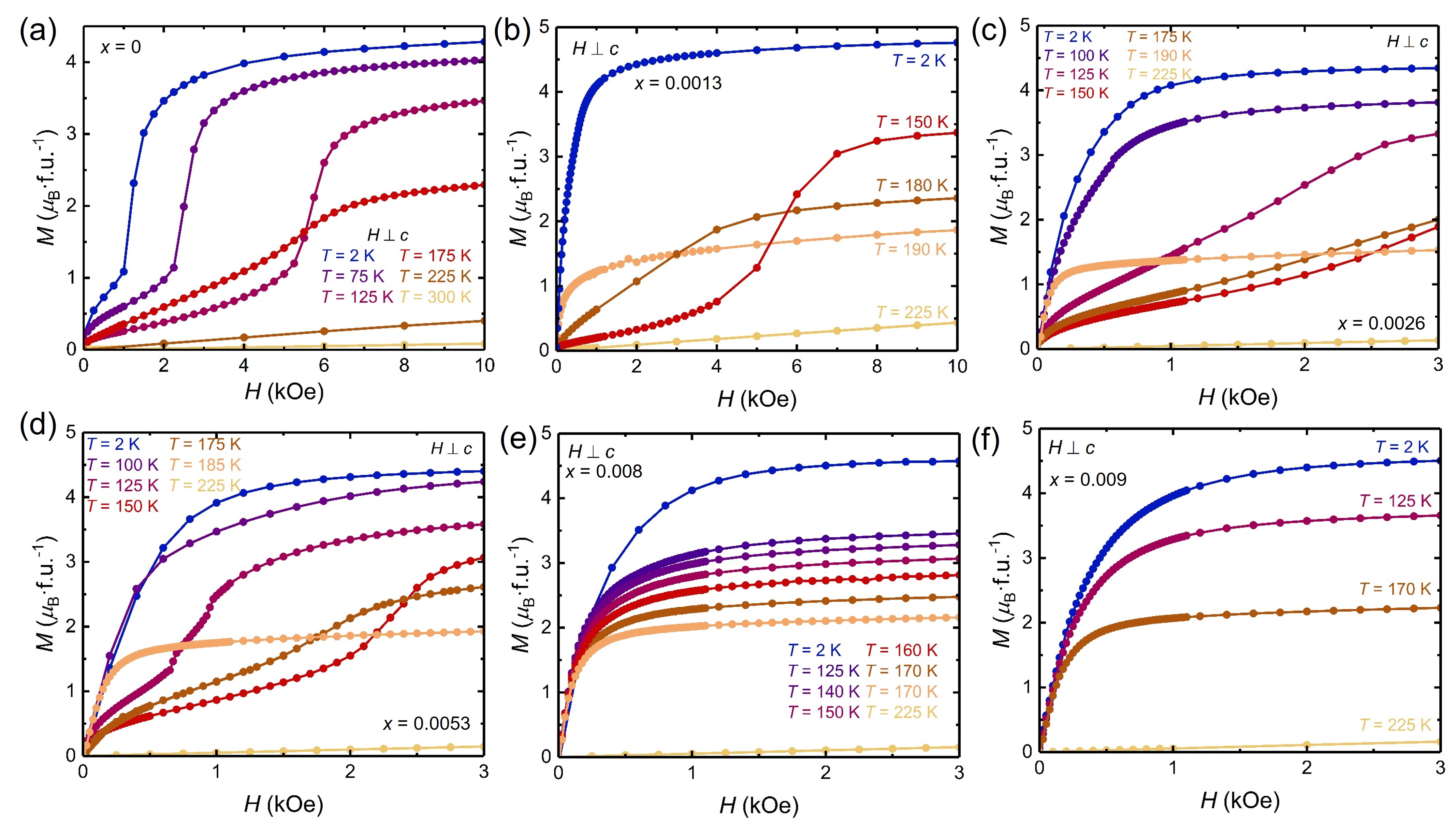}
    \caption[]{Field-dependent magnetization isotherms measured at temperatures corresponding to different parts of the phase diagram for (a) \textit{x} = 0, (b) \textit{x} = 0.0013, (c) \textit{x} = 0.0026, (d) \textit{x} = 0.0053, (e) \textit{x} = 0.008, (f) \textit{x} = 0.009. Note that the \textit{x}-axis scale for (a) and (b) extends to 10 kOe to observe the metamagnetic transitions.}
    \label{MH_lowx}
\end{figure*}

\subsection{Magnetic and transport properties of Mn(Pt$_{1-x}$Pd$_x$)$_5$P} MnPt$_5$P enters into a spin-canted antiferromagnetic state at \textit{T}$_N$ $\approx$ 190 K,\autocite{gui2020chemical, SladeCanfieldZAAC} and preliminary data collected on MnPd$_5$P indicated that this material becomes ferromagnetic near room temperature.\autocite{SladeCanfieldZAAC} To understand how the magnetic state evolves as Pt is replaced with Pd, we conducted temperature and field dependent magnetization and transport measurements on our Mn(Pt$_{1-x}$Pd$_x$)$_5$P samples, and the results are summarized in the temperature-composition (\textit{T}–\textit{x}) phase diagram given in Figure \ref{T-x}. The magnetization and transport data are outlined in Figures \ref{MT} and \ref{RT}, and Figure \ref{MH_lowx} shows the field dependent magnetization isotherms collected at salient temperatures for \textit{x} $<$ 0.010 samples. 

Temperature dependent magnetization (\textit{M/H}) collected at \textit{H} = 1 kOe with the field applied within the easy ab-plane (\textit{H} $\perp$ \textit{c}) for the Mn(Pt$_{1-x}$Pd$_x$)$_5$P single crystals is presented in Figure \ref{MT}a, and Figure \ref{MT}b shows \textit{H} = 50 Oe data for \textit{x} $<$ 0.010 samples (See Figure $\ref{MH_highx}$ in the Appendix for the anisotropic \textit{M}(\textit{H}) results). At low Pd substitution (\textit{x} $<$ 0.010), the initially narrow, antiferromagnetic-like, peak observed at 192 K for pure MnPt$_5$P substantially broadens as \textit{x} increases and forms a plateau-like maxima centered near $\approx$ 180 K, whereas the weak upturn in \textit{M/H} below 100 K in MnPt$_5$P becomes a sharp increase reminiscent of ferromagnetic ordering. The ferromagnetic-like upturn moves to higher temperatures as \textit{x} increases, eventually merging with the initial higher temperature transition such that by \textit{x} = 0.008, the \textit{M/H} data shows a very rapid increases at $\approx$ 193 K followed by a second, subtle, increase beginning at $\approx$150 K. Above \textit{x} $>$ 0.010, only a single ferromagnetic transition is observed, and the Curie temperature (determined more precisely from resistance measurements discussed below) increases with Pd alloying to a maximum at 312 K for \textit{x} = 0.62 before falling gradually back to $\approx$ 295 K for pure MnPd$_5$P.

To complement the temperature dependent \textit{M/H} data, we also measured the resistance of each sample from 1.8–375 K and present the data in Figure \ref{RT}. The \textit{R}(\textit{T}) results for low \textit{x} samples are given in Figure \ref{RT}a and the derivatives d\textit{R}/d\textit{T} used to assign the transition temperatures are shown in Figure \ref{RT}b.\autocite{PhysRevLett.20.665}  Corresponding \textit{R}(\textit{T}) and d\textit{R}/d\textit{T} data for the \textit{x} $>$ 0.03 samples is shown in Figures \ref{RT}c and \ref{RT}d. As expected, all samples have metallic resistance that decreases with cooling, and the \textit{R}(\textit{T}) datasets each show a clear kink followed by a rapid drop at the initial magnetic transition temperature \textit{T}$_1$, characteristic of losing spin-disorder scattering as the samples enter a magnetically ordered state. The residual resistance ratios, \textit{RRR} = \textit{R}(375 K)/\textit{R}(1.8 K), are shown in the appendix in Figure \ref{RT_hysteresis_RRR}b and are minimized at \textit{x} = 0.54, consistent with the expectation for stronger scattering associated with the crystallograpically disordered Pt and Pd atoms in the alloys. 

At low Pd fraction, the samples with 0.0013 $\leq$ \textit{x} $\leq$ 0.0053 show a second transition \textit{T}$_2$ just below the first, which is suppressed from $\approx$187 K at \textit{x} = 0.0013 to $\approx$177 K at \textit{x} = 0.0053. Further cooling reveals another lower temperature transition \textit{T}$_3$ that increases with Pd alloying from $\approx$71 K for \textit{x} = 0.00125 to 160 K at \textit{x} = 0.009. We note that the signature of \textit{T}$_2$ is lost in the \textit{x} = 0.008 and \textit{x} = 0.009 resistance data, which is likely due to the close proximity of \textit{T}$_2$ and \textit{T}$_3$ at these compositions. This is consistent with the \textit{M/H} data (see Figure \ref{MT}b), which shows the lower temperature ferromagnetic feature (\textit{T}$_3$) essentially merging with the higher temperature transitions. \textit{T}$_3$ is hysteretic between warming and cooling, implying it is first order (see Figure \ref{RT_hysteresis_RRR} in the Appendix for a close up view). The resistance curves for 0.033 $\leq$ \textit{x} $\leq$ 1 (Figure 5b) only show a single transition that increases rapidly with \textit{x} and is maximized at $\approx$ 312 K before falling back to 295 K for pure MnPd$_5$P, consistent with the \textit{M/H} data in Figure \ref{MT}.

Using the transitions identified in the temperature dependent \textit{M/H} and \textit{R}(\textit{T}) data discussed above, we can identify the unique regions of the \textit{T}–\textit{x} phase diagram shown in Figure \ref{T-x}. We find that the initial transition at 192 K in pure MnPt$_5$P splits into two transitions, \textit{T}$_1$ and \textit{T}$_2$, upon even minute, almost homeopathic, levels of Pd substitution. \textit{T}$_1$ appears to be ferromagnetic and increases with \textit{x} to a maximum at 321 K for \textit{x} = 0.62, whereas \textit{T}$_2$ decreases gradually with \textit{x}. The low-\textit{x} samples show a final third transition \textit{T}$_3$ that increases with \textit{x} and intersects \textit{T}$_2$ at approximately \textit{x} $\approx$ 0.009, such that \textit{T}$_2$ and \textit{T}$_3$ delineate a “bubble phase” on the phase diagram that extends out to \textit{x} $\approx$ 0.009. Below the bubble, the low-\textit{x} samples recover the original ferromagnetic state entered upon cooling through \textit{T}$_1$.

To better determine the type of order found in each region of the phase diagram, Figure \ref{MH_lowx} presents magnetization isotherms measured at salient temperatures for the \textit{x} $<$ 0.010 Mn(Pt$_{1-x}$Pd$_x$)$_5$P samples. (Figure \ref{MH_highx} shows the \textit{M}(\textit{H}) data for higher \textit{x} samples over a much higher applied field range.)  As shown in Figure \ref{MH_lowx}a, the \textit{M}(\textit{H}) curves for pure MnPt$_5$P show a series of metamagnetic transitions that shift to lower field as the temperature is lowered, indicating antiferromagnetic order. As outlined in our prior work,\autocite{SladeCanfieldZAAC} the magnetization curves for MnPt$_5$P all show a small saturation and measurable hysteresis at the lowest fields (under 0.2 kOe), implying that the antiferromagnetic order also has a small ferromagnetic (\textit{q} = 0) component (we use * in Figure \ref{T-x}b to denote the ferromagnetic component to the otherwise primarily antiferromagnetic state). Importantly, despite having a broad upturn in \textit{M/H} below $\approx$ 100 K (see Figures \ref{MT}a and \ref{MT}b), the 2 K \textit{M}(\textit{H}) isotherm for MnPt$_5$P is qualitatively the same as the higher temperature datasets, with a clear metamagnetic transition observed at $\approx$ 1 kOe, indicating that MnPt$_5$P remains antiferromagnetic down to at least 2 K, which is consistent with the \textit{R}(\textit{T}) data for MnPt$_5$P showing only a single 192 K transition.  This said, the very low metamagnetic field, that decreases with decreasing temperature, rather than the more standard increasing with decreasing temperature, strongly suggests a close energetic proximity to an ordered state with a larger ferromagnetic component.

Upon introduction of Pd, the \textit{M}(\textit{H}) isotherms measured above the “bubble” region of the phase diagram, at temperatures between \textit{T}$_1$ and \textit{T}$_2$, show a swift rise in \textit{M} at low field followed by saturation at above $\approx$ 1 kOe, implying that \textit{T}$_1$ is a ferromagnetic transition (see 190 K data in Figures \ref{MH_lowx}b–\ref{MH_lowx}c and 185 K data in Figure \ref{MH_lowx}d). Within the bubble, between \textit{T}$_2$ and \textit{T}$_3$, the \textit{M}(\textit{H}) curves all show metamagnetic transitions that generally move to lower fields as the temperature is lowered (for a given value of \textit{x}). Likewise, the \textit{M}(\textit{H}) datasets all show a small but measurable low-field saturation below $\approx$0.2 kOe. Together, this information suggests that the bubble phase is the same spin-canted AFM* state found in pure MnPt$_5$P. Below \textit{T}$_3$, the \textit{M}(\textit{H}) isotherms for Pd containing samples again are characteristic ferromagnetic behavior, with a rapid increase in \textit{M} at low field followed by saturation at $\approx$ 4.5 $\mu_{\text{B}}$/f.u. When the Pd fraction rises above \textit{x} $\approx$ 0.008-0.009, the \textit{M}(\textit{H}) isotherms show classic easy-plane ferromagnetic behavior at all temperatures below \textit{T}$_1$ (see Figure \ref{MH_highx} in the Appendix for \textit{M}(\textit{H}) data for \textit{x} $>$ 0.010).

The \textit{M}(\textit{H}) data collected in the ferromagnetic phase (below \textit{T}$_3$ for \textit{x} $<$ 0.010 and below \textit{T}$_1$ for \textit{x} $>$ 0.010) show very small ($\approx$ 10-20 Oe), almost negligible, hysteresis on raising and lowering the field, implying that the ferromagnetism in Mn(Pt$_{1-x}$Pd$_x$)$_5$P single crystals is very soft. Moreover, the \textit{M}(\textit{H}) results show that the Mn(Pt$_{1-x}$Pd$_x$)$_5$P samples have relatively strong magnetic anisotropy where the \textit{ab}-plane is the easy direction. Anisotropy fields \textit{H}$_{\text{A}}$ estimated by extrapolating the tangents of the \textit{H} $\perp$ \textit{c} and \textit{H} $\parallel$ \textit{c} datasets decrease monotonically as the Pd fraction rises (see Figure \ref{MH_highx} in the Appendix), from $\approx$ 108 kOe for \textit{x} = 0.022 Pd to $\approx$ 10 kOe in MnPd$_5$P, which likely reflects the decreasing strength of spin orbit coupling accompanying substitution of the Pt with the smaller \textit{Z} Pd. 

\subsection{Discussion}

\begin{figure}[!t]
    \centering
    \includegraphics[width=\linewidth]{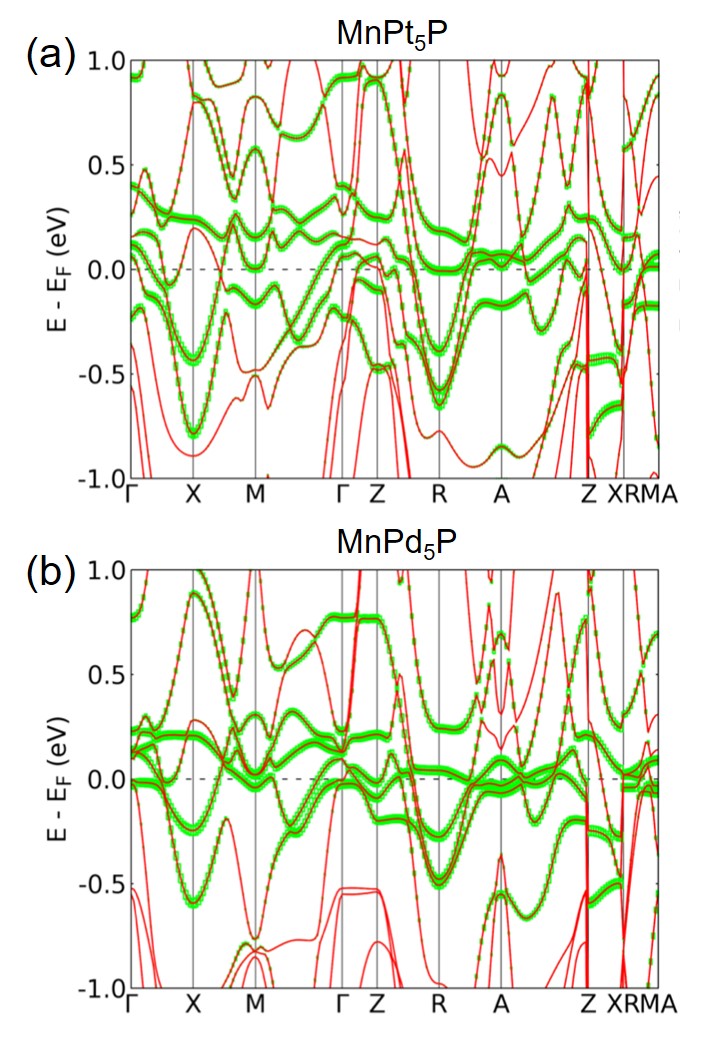}
    \caption[]{Electronic band structures calculated for (a) MnPt$_5$P and (b) MnPd$_5$P. The green shading represents the relative projection of Mn-3\textit{d} orbitals to the electronic bands.}
    \label{BandStructure}
\end{figure}

Our magnetic and transport measurements strongly suggest that the energy difference between ferromagnetic and antiferromagnetic states in MnPt$_5$P is exceptionally small, such that even the small perturbation of \textit{x} $\approx$ 0.02 Pd substitution on the Pt sites is sufficient to stabilize purely ferromagnetic order. At very low, essentially homeopathic levels of Pd \textit{x} $<$ 0.01, both antiferromagnetic and ferromagnetic phases are observed, and the antiferromagnetic state forms a bubble phase spanning pure MnPt$_5$P to \textit{x} $\approx$ 0.008-0.009. Whereas this magnetic phase diagram, shown in figure \ref{T-x}b, describes the transitions below 200 K, non-monotonic changes in the \textit{a}-lattice parameter and unit cell volume are also possibly detected at room temperature over essentially the same range of low \textit{x} (see Figure \ref{PXRD}b and \ref{PXRD}c). This suggests that MnPt$_5$P undergoes a transition, with very small Pd substitution, that manifests itself both in the lattice as well as the nature of the magnetic interactions. A clear candidate would be a Lifshitz type transition where the Fermi surface topology changes (i.e. small pockets appear/disappear) with small Pd substitution. Such an electronic transition can lead to changes in the density of states (DOS) at the Fermi energy as well as changes in the generalized electronic susceptibility, $\chi$(\textit{q}), which governs whether the magnetic order is anti- or ferromagnetic. To explore this possibility, we calculated electronic band structures for paramagnetic MnPt$_5$P and MnPd$_5$P.

The band structures are displayed in Figure \ref{BandStructure}, where the green shading represents the projection of Mn-3\textit{d} orbitals to the electronic states. As expected, the calculations indicate both MnPt$_5$P and MnPd$_5$P are metals, with multiple well dispersed bands crossing the Fermi energy (\textit{E}$_{\text{F}}$). In both compounds, most of the bands near \textit{E}$_{\text{F}}$ are composed of Mn-3\textit{d} based states. Most importantly, the band structure for pure MnPt$_5$P has several bands that graze, or come very close to \textit{E}$_{\text{F}}$ at the \textit{M}, \textit{R}, and \textit{A} points in the Brillouin zone, as well as a Dirac-like set of bands along \textit{X}--\textit{M}. In MnPd$_5$P, the flat band sections near the \textit{M} and \textit{R} points move above \textit{E}$_{\text{F}}$, and the band along the \textit{X--M} direction also moves up in energy. Furthermore, new flat band sections appear near \textit{E}$_{\text{F}}$ in MnPd$_5$P along the $\Gamma$--\textit{X} and \textit{A}--\textit{Z} directions. Admittedly, comparison of the end members MnPt$_5$P and MnPd$_5$P represents an extreme perturbation in reference to the \textit{x} $\approx$ 0.02 needed to stabalize purely ferromagnetic order in Mn(Pt$_{5-x}$Pd$_x$)$_5$P; however, the calculations do show that MnPt$_5$P has multiple pockets very near \textit{E}$_{\text{F}}$ that are substantially changed in MnPd$_5$P, suggesting a Lifshitz transition is at least plausible in Mn(Pt$_{5-x}$Pd$_x$)$_5$P. Subsequent measurements that directly probe the electronic states at \textit{E}$_{\text{F}}$, such as the thermopower and Hall effect, would be needed to explore and test this proposal.  

\section{Summary and Conclusions}

We determined that ferromagnetic MnPd$_5$P adopts the anti-CeCoIn$_5$ structure with the space group \textit{P}4/\textit{mmm} and conducted a detailed substitutional study with its isostructural antiferromagnetic analogue MnPt$_5$P. We demonstrate a solution route to grow large single crystals of both MnPd$_5$P and the alloys Mn(Pt$_{1-x}$Pd$_x$)$_5$P. EDS and X-ray diffraction data support the formation of a full Mn(Pt$_{1-x}$Pd$_x$)$_5$P solid solution that maintains the tetragonal anti-CeCoIn$_5$ structure. The magnetic data show that the primarily antiferromagnetic state in pure MnPt$_5$P is extremely sensitive to Pd substitution, and as little as \textit{x} $>$ 0.010 Pd stabilizes purely ferromagnetic order. At low \textit{x} $<$ 0.010, the single antiferromagnetic transition in MnPt$_5$P splits into a higher temperature ferromagnetic transition and lower temperature ferromagnetic-to-antiferromagnetic and lower temperature antiferromagnetic to ferromagnetic transition.  The antiferromagnetic region forms a bubble-region in the \textit{T--x} phase diagram which persists up to \textit{x} $\approx$ 0.008–0.009, and further cooling recovers the original ferromagnetic state as the samples approach base temperature. Room temperature values of the \textit{a}-lattice parameter and unit-cell-volume also manifest anomalous behavior for \textit{x} $<$ $\approx$ 0.010, suggesting that some electronic topological phase transition, such as a Lifshitz transition, may be responsible for the changes in both magnetic ordering as well as structural features. Electronic band structure calculations indicate pure MnPt$_5$P has several pockets close to the Fermi level that could be involved in such a transition. Beyond \textit{x} $>$ 0.010, the ferromagnetic Curie temperature is substantially enhanced with Pd incorporation to $\approx$ 312 K at \textit{x} = 0.62 before falling back to 295 K in pure MnPd$_5$P. All Mn(Pt$_{1-x}$Pd$_x$)$_5$P samples have strong magnetic anisotropy in which the \textit{ab}-plane is the easy direction, and the anisotropy field decreases from $\approx$108 kOe for \textit{x} = 0.022 Pd to $\approx$10 kOe for MnPd$_5$P, likely a result of reduced spin orbit coupling.  

\section{Acknowledgements}

Work at Ames National Laboratory (T.J.S., N.F., T.R.S, J.S., L.L.W., S.L.B., P.C.C.) were supported by the U.S. Department of Energy, Office of Science, Basic Energy Sciences, Materials Sciences and Engineering Division. Ames National Laboratory is operated for the U.S. Department of Energy by Iowa State
University under Contract No. DE-AC02-07CH11358. T.J.S., P.C.C., and L.L.W. were supported by the Center for Advancement of Topological Semimetals (CATS), an Energy Frontier Research Center funded by the U.S. Department of Energy Office of Science, Office of Basic Energy Sciences, through the Ames National Laboratory under its Contract No. DE-AC02-07CH11358 with Iowa State University. The work at Rutgers is supported by Beckman Young Investigator award and NSF-DMR-2053287. C. -J.K. and G.K. were supported by the U.S. Department of Energy, Office of Science (Basic Energy Science) as a part of the Computational Materials Science Program through the Center for Computational Design of Functional Strongly Correlated Materials and Theoretical Spectroscopy under DOE grant no. DE-FOA-0001276. C.-J.K. also acknowledges support by NRF grant No. 2022R1C1C1008200. A portion of this work was performed at the National High Magnetic Field Laboratory, which is supported by National Science Foundation Cooperative Agreement No. DMR-1644779 and the State of Florida.

\vskip 0.25cm
\noindent
*corresponding authors' email: slade@ameslab.gov, canfield@ameslab.gov

\vskip 0.25cm
\noindent
$^{||}$T.J.S. and R.S.D.M. are equally contributing authors.

\vskip 0.25cm
\noindent
\textbf{\textit{Conflicts of Interest}}

The authors have no conflicts of interest to declare.

\section{Appendix}

\subsection{EDS analysis of Mn(Pt$_{1-x}$Pd$_x$)$_5$P}

\begin{table}[!b]
  \small
  \centering
  \caption{Chemical compositions determined from EDS analysis for the solution-grown single crystals of Mn(Pt$_{1-x}$Pd$_x$)$_5$P. The nominal compositions used for the growth were Mn$_9$Pt$_{71-y}$Pd$_y$P$_{20}$, and we also give the nominal Pd:Pt fraction (\textit{x}) in each. The EDS values of \textit{x} represent the averages of 3-6 scans on each sample and the error bars were obtained considering both the EDS fitting errors and standard deviations of each measurement (see text).}
    \resizebox{\linewidth}{!}{\begin{tabular}{ccc}
    \toprule
    \multicolumn{1}{l}{Nominal \textit{y}} & \multicolumn{1}{l}{Nominal \textit{x}} & \multicolumn{1}{l}{EDS \textit{x}} Mn(Pt$_{1-x}$Pd$_x$)$_5$P \\
    0.25 & 0.0035 & 0.00013 $\pm$ 0.001 \\
    0.5 & 0.0070 & 0.003 $\pm$ 0.001 \\
    1 & 0.014 & 0.0053 $\pm$ 0.0005 \\
    1.5 & 0.021 & 0.0082 $\pm$ 0.0007 \\
    1.75 & 0.025 & 0.009 $\pm$ 0.001 \\
    2.25 & 0.032 & 0.013 $\pm$ 0.002 \\
    3  & 0.042 & 0.022 $\pm$ 0.001 \\
    6  & 0.0845 & 0.033 $\pm$ 0.003 \\
    9  & 0.13 & 0.0635 $\pm$ 0.003 \\
    16 & 0.225 & 0.116 $\pm$ 0.005 \\
    23 & 0.32 & 0.188 $\pm$ 0.005 \\
    47 & 0.66 & 0.543 $\pm$ 0.004 \\
    50 & 0.70 & 0.623 $\pm$ 0.004 \\
    \bottomrule
    \end{tabular}}%
  \label{EDS_SC}%
\end{table}%

\begin{table}[!b]
  \small
  \centering
  \caption{Chemical compositions determined from EDS analysis for the polycrystalline samples of Mn(Pt$_{1-x}$Pd$_x$)$_5$P. The nominal compositions used for the growth were Mn(Pt$_{1-y}$Pd$_y$)$_5$P.}
    \resizebox{\linewidth}{!}{\begin{tabular}{cc}
    \toprule
    \multicolumn{1}{l}{Nominal \textit{x} Mn(Pt$_{1-x}$Pd$_x$)$_5$P} & EDS \textit{x} Mn(Pt$_{1-x}$Pd$_x$)$_5$P \\
    \midrule
    0.2 & 0.188(5) \\
    0.4 & 0.39(2) \\
    0.5   & 0.502(5) \\
    0.6 & 0.589(3) \\
    0.8 & 0.81(1) \\
    \bottomrule
    \end{tabular}}%
  \label{EDS_PC}%
\end{table}%

\begin{table*}[!t]
  \centering
  \caption{Single crystal structure refinement information for Mn(Pt$_{1-x}$Pd$_x$)$_5$P at 300(2) K. (The standard deviations are indicated by the values in parentheses). The single crystals used for data collection and refinement were picked from the sintered pellets.}
    \resizebox{\linewidth}{!}{\begin{tabular}{rrrrrr}
    \toprule
    Loaded composition & MnPd$_5$P & Mn(Pt$_{0.2}$Pd$_{0.8}$)$_5$P & Mn(Pt$_{0.4}$Pd$_{0.6}$)$_5$P & Mn(Pt$_{0.5}$Pd$_{0.5}$)$_5$P & Mn(Pt$_{0.8}$Pt$_{0.2}$)$_5$P \\
    Refined Formula & MnPd$_5$P & Mn(Pt$_{0.172(4)}$Pd$_{0.828(4)}$)$_5$P & Mn(Pt$_{0.454}$Pd$_{0.546}$)$_5$P & Mn(Pt$_{0.48}$Pd$_{0.52}$)$_5$P & Mn(Pt$_{0.58}$Pd$_{0.42}$)$_5$P \\
    F.W. (g/mol) & \multicolumn{1}{r}{617.91} & \multicolumn{1}{r}{694.18} & \multicolumn{1}{r}{819.24} & \multicolumn{1}{r}{848.5} & \multicolumn{1}{r}{875.11} \\
    Space group; \textit{Z} & \textit{P}4/\textit{mmm}; 1 & \textit{P}4/\textit{mmm}; 1 & \textit{P}4/\textit{mmm}; 1 & \textit{P}4/\textit{mmm}; 1 & \textit{P}4/\textit{mmm}; 1 \\
    \textit{a}(\AA) & 3.899(2) & 3.894(1) & 3.887(2) & 3.888 (2) & 3.901(3) \\
    \textit{c}(\AA) & 6.867(4) & 6.855(1) & 6.853(2) & 6.861(2) & 6.892(4) \\
    \textit{V} (\AA$^3$) & 104.42(2) & 103.93(2) & 103.54(4) & 103.73(5) & 104.93(11) \\
    $\theta$ range (º) & 2.966-34.770 & 5.236-34.828 & 5.246-34.892 & 2.969-3.874 & 5.919-34.646 \\
    No. reflections; \textit{R}$_{\text{int}}$ & 578; 0.0609 & 1397; 0.0293 & 999; 0.0444 & 904; 0.0549 & 214; 0.0233 \\
    No. independent reflections & \multicolumn{1}{r}{170} & \multicolumn{1}{r}{173} & \multicolumn{1}{r}{166} & \multicolumn{1}{r}{167} & \multicolumn{1}{r}{94} \\
    No. parameters & \multicolumn{1}{r}{12} & \multicolumn{1}{r}{14} & \multicolumn{1}{r}{14} & \multicolumn{1}{r}{14} & \multicolumn{1}{r}{14} \\
    \textit{R}$_{\text{1}}$: w\textit{R}$_{\text{2}}$ (\textit{I} $>$ 2(\text{I})) & 0.0509; 0.1204 & 0.0253; 0.0603 & 0.0429; 0.1026 & 0.0417; 0.1080 & 0.0373; 0.0947 \\
    Goodness of fit & \multicolumn{1}{r}{1.282} & \multicolumn{1}{r}{1.346} & \multicolumn{1}{r}{1.331} & \multicolumn{1}{r}{1.429} & \multicolumn{1}{r}{1.154} \\
    Diffraction peak and hole (\textit{e}$^-$/\AA$^3$) & 2.656; -1.863 & 2.632; -1.926 & 7.771; -3.765 & 7.323; 5.219 & 2.579; -3.540 \\
    Temperature (K) & 300 (2) & 299 (2) & 301 (2) & 300 (2) & 301 (2) \\
    \bottomrule
    \end{tabular}}%
  \label{Mn(Pt1-xPdx)5P_crystallography}%
\end{table*}%

\begin{table}[!b]
  \centering
  \caption{Atomic coordinates, occupancies and isotropic displacement parameters of Mn(Pt$_{1-x}$Pd$_x$)$_5$P at 300(2) K. (\textit{U}$_{\text{eq}}$ is defined as one-third of the trace of the orthogonalized \textit{U}$_{ij}$ tensor (\AA$^2$)).}
    \resizebox{\linewidth}{!}{\begin{tabular}{rrrrrrr}
    \toprule
    Atom & Wyckoff. & \multicolumn{1}{l}{Occ.} & \multicolumn{1}{l}{\textit{x}} & \multicolumn{1}{l}{\textit{y}} & \textit{z}   & \textit{U}$_{\text{eq}}$ \\
    \midrule
    \multicolumn{7}{c}{MnPd$_5$P} \\
    \midrule
    Pd1 & 4i  & 1   & 0   & \multicolumn{1}{l}{½} & 0.2948 (1) & 0.015(1) \\
    Pd2 & 1a  & 1   & 0   & 0   & \multicolumn{1}{r}{0} & 0.012(2) \\
    Mn3 & 1c  & 1   & \multicolumn{1}{l}{½} & \multicolumn{1}{l}{½} & \multicolumn{1}{r}{0} & 0.022(1) \\
    P4  & 1b  & 1   & 0   & 0   & ½   &  0.016(2) \\
    \midrule
    \multicolumn{7}{c}{Mn(Pt$_{0.2}$Pd$_{0.8}$)$_5$P} \\
    \midrule
    Pd1 & 4i  & \multicolumn{1}{l}{0.84(2)} & 0   & \multicolumn{1}{l}{½} & 0.29385(9)  & 0.0058(2) \\
    Pt2 & 4i  & \multicolumn{1}{l}{0.22(2)} & 0   & \multicolumn{1}{l}{½} & 0.2948(1) & 0.0058(2) \\
    Pd3 & 1a  & \multicolumn{1}{l}{0.78(2)} & 0   & 0   & \multicolumn{1}{r}{0} & 0.0046(3) \\
    Pt4 & 1a  & \multicolumn{1}{l}{0.22(2)} & 0   & 0   & \multicolumn{1}{r}{0} & 0.0046(3) \\
    Mn3 & 1c  & 1   & \multicolumn{1}{l}{½} & \multicolumn{1}{l}{½} & \multicolumn{1}{r}{0} & 0.0128(9) \\
    P4  & 1b  & 1   & 0   & 0   & ½   & 0.0076(11) \\
    \midrule
    \multicolumn{7}{c}{Mn(Pt$_{0.4}$Pd$_{0.6}$)$_5$P} \\
    \midrule
    Pd1 & 4i  & \multicolumn{1}{l}{0.57(6)} & 0   & \multicolumn{1}{l}{½} & 0.29296(17)  & 0.0056(4) \\
    Pt2 & 4i  & \multicolumn{1}{l}{0.43(6)} & 0   & \multicolumn{1}{l}{½} & 0.29296(17) & 0.0056(4) \\
    Pd3 & 1a  & \multicolumn{1}{l}{0.55(6)} & 0   & 0   & \multicolumn{1}{r}{0} & 0.0037(6) \\
    Pt4 & 1a  & \multicolumn{1}{l}{0.45(6)} & 0   & 0   & \multicolumn{1}{r}{0} & 0.0037(6) \\
    Mn3 & 1c  & 1   & \multicolumn{1}{l}{½} & \multicolumn{1}{l}{½} & \multicolumn{1}{r}{0} & 0.020(2) \\
    P4  & 1b  & 1   & 0   & 0   & ½   &  0.006(2) \\
    \midrule
    \multicolumn{7}{c}{Mn(Pt$_{0.5}$Pd$_{0.5}$)$_5$P} \\
    \midrule
    Pd1 & 4i  & \multicolumn{1}{l}{0.50(8)} & 0   & \multicolumn{1}{l}{½} & 0.29274(18)  & 0.0078(5) \\
    Pt2 & 4i  & \multicolumn{1}{l}{0.50(8)} & 0   & \multicolumn{1}{l}{½} & 0.29274(18) & 0.0078(5) \\
    Pd3 & 1a  & \multicolumn{1}{l}{0.42(6)} & 0   & 0   & \multicolumn{1}{r}{0} & 0.0049(7) \\
    Pt4 & 1a  & \multicolumn{1}{l}{0.58(6)} & 0   & 0   & \multicolumn{1}{r}{0} & 0.0049(7) \\
    Mn3 & 1c  & 1   & \multicolumn{1}{l}{½} & \multicolumn{1}{l}{½} & \multicolumn{1}{r}{0} & 0.015(3) \\
    P4  & 1b  & 1   & 0   & 0   & ½   &  0.011(3) \\
    \midrule
    \multicolumn{7}{c}{Mn(Pt$_{0.6}$Pd$_{0.4}$)$_5$P} \\
    \midrule
    Pd1 & 4i  & \multicolumn{1}{l}{0.43(6)} & 0   & \multicolumn{1}{l}{½} & 0.29254(11)  & 0.0084(6) \\
    Pt2 & 4i  & \multicolumn{1}{l}{0.57(6)} & 0   & \multicolumn{1}{l}{½} & 0.29254(11) & 0.0084(6) \\
    Pd3 & 1a  & \multicolumn{1}{l}{0.38(6)} & 0   & 0   & \multicolumn{1}{r}{0} & 0.0068(7) \\
    Pt4 & 1a  & \multicolumn{1}{l}{0.62(6)} & 0   & 0   & \multicolumn{1}{r}{0} & 0.0068(7) \\
    Mn3 & 1c  & 1   & \multicolumn{1}{l}{½} & \multicolumn{1}{l}{½} & \multicolumn{1}{r}{0} & 0.012(3) \\
    P4  & 1b  & 1   & 0   & 0   & ½   & 0.014(4) \\
    \bottomrule
    \end{tabular}}%
  \label{Mn(Pt1-xPdx)P_refineatoms}%
\end{table}%

\begin{figure}[!t]
    \centering
    \includegraphics[width=\linewidth]{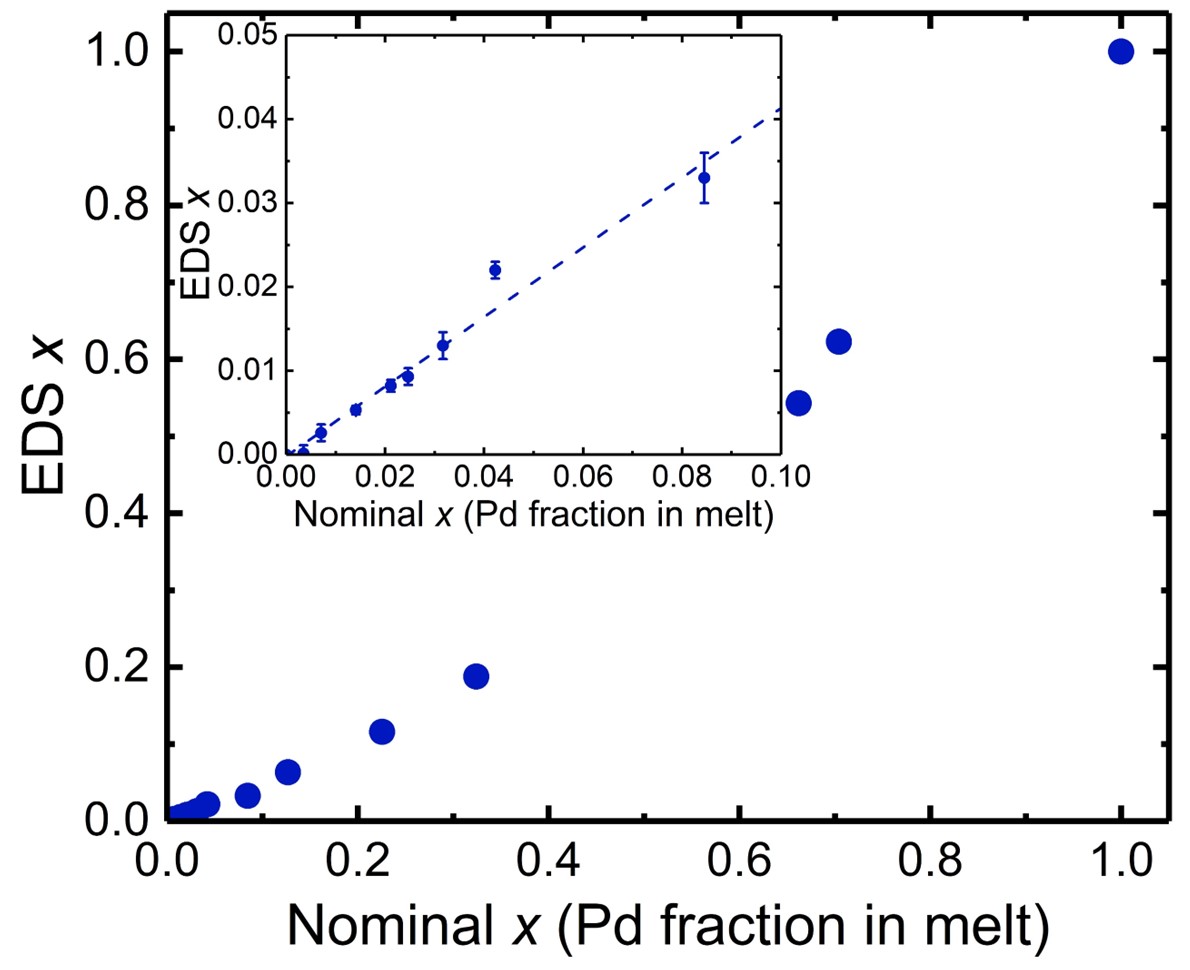}
    \caption[]{EDS values of \textit{x} Pd in Mn(Pt$_{1-x}$Pd$_x$)$_5$P compared to the nominal \textit{x} added to the starting crystal growth compositions. The inset shows a closeup of the low-\textit{x} data, and the dashed line is a linear fit to the data up to nominally \textit{x} = 0.0845.}
    \label{xnomxEDS}
\end{figure}

\begin{figure*}[!t]
    \centering
    \includegraphics[width=\linewidth]{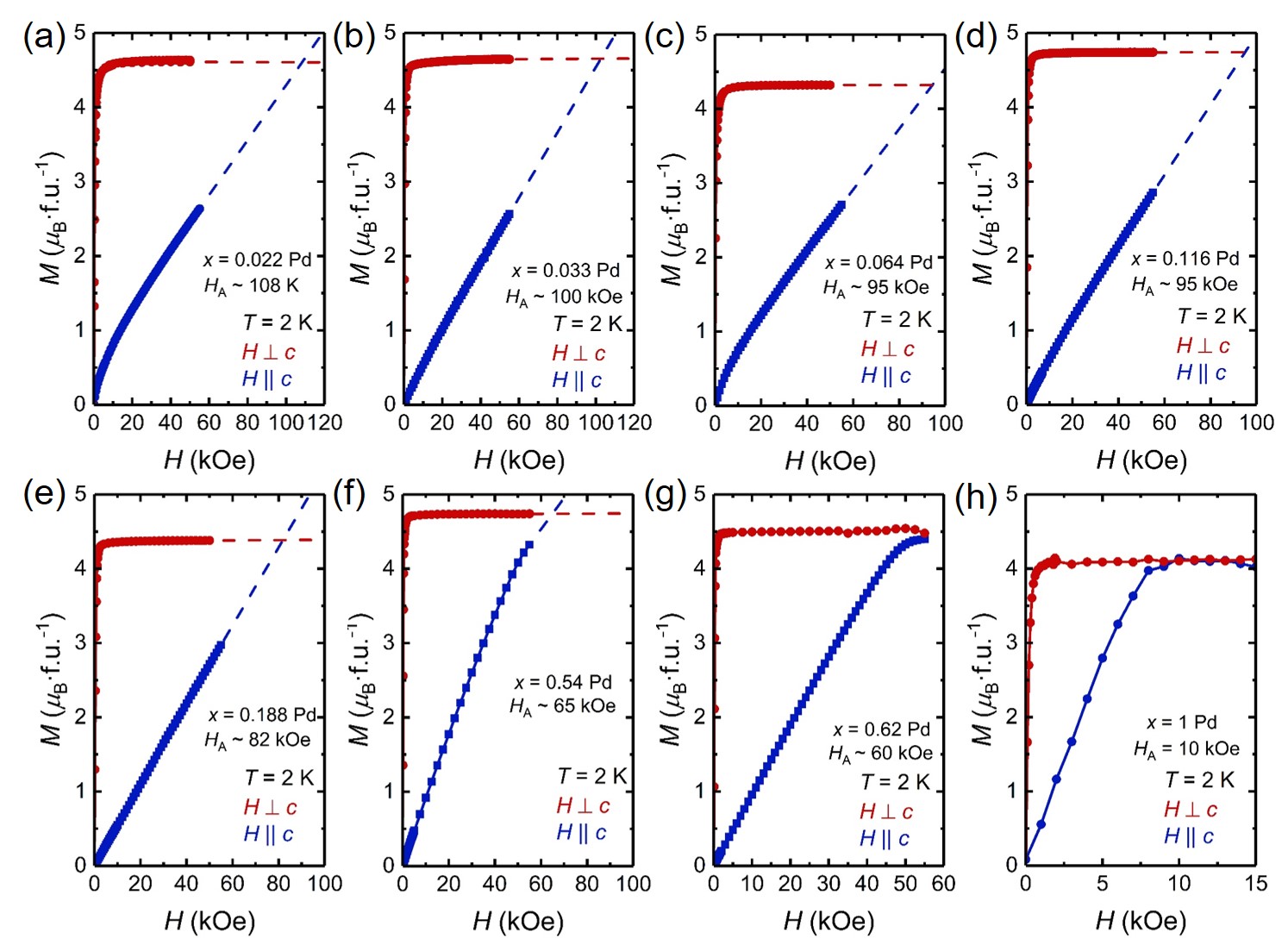}
    \caption[]{Field dependent magnetization isotherms measured at 2 K for all Pd containing Mn(Pt$_{1-x}$Pd$_x$)$_5$P (0.022 $\geq$ \textit{x} $\geq$ 1) single crystals. The small degree of non-linearity observed at low fields for \textit{x} = 0.022 and \textit{x} = 0.064 Pd likely is from small misorientation of the sample. The dotted lines show the extrapolation of the tangents used to estimate the anisotropy fields \textit{H}$_{\text{A}}$.}
    \label{MH_highx}
\end{figure*}

\begin{figure}[!t]
    \centering
    \includegraphics[width=\linewidth]{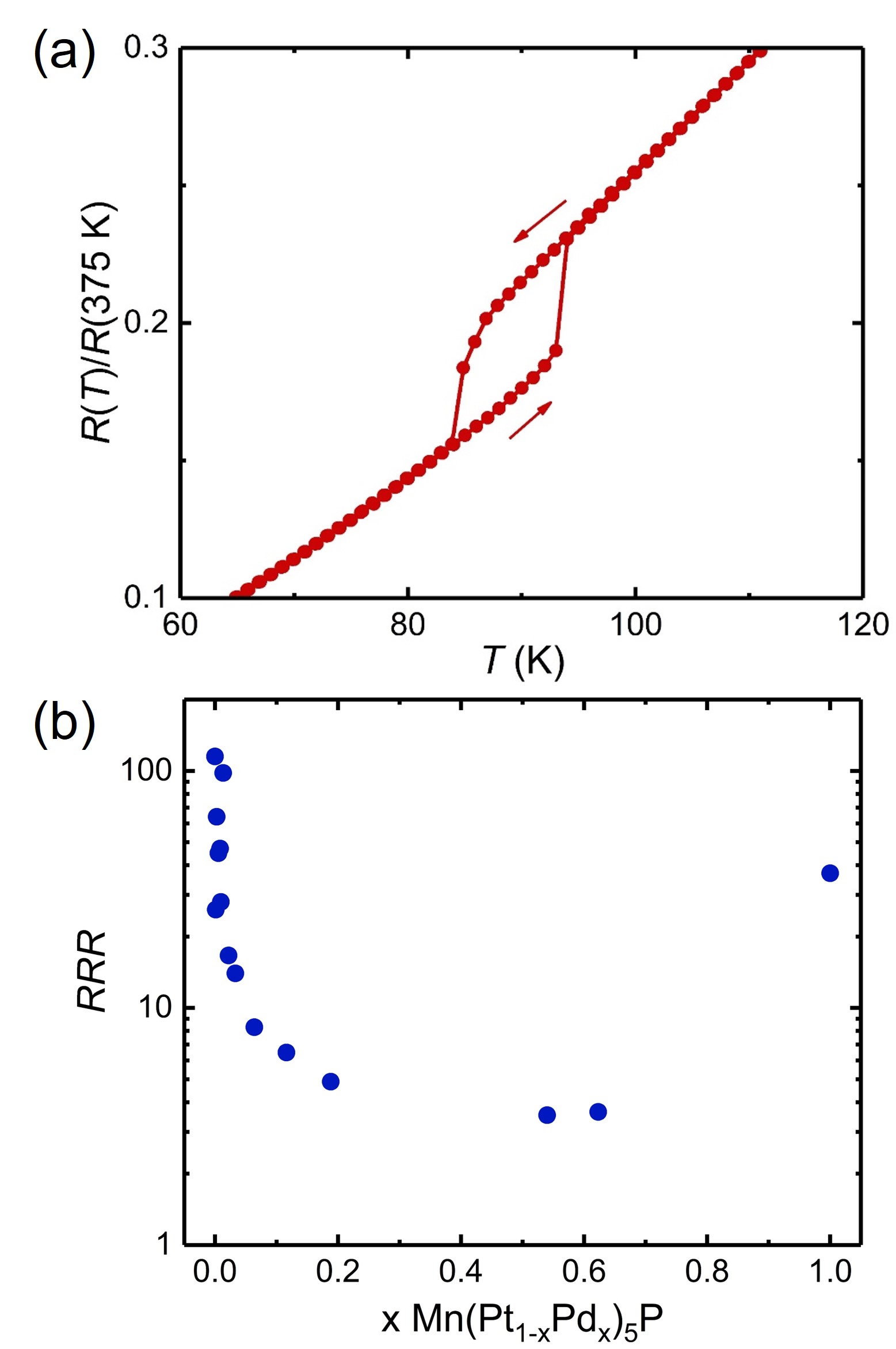}
    \caption[]{(a) Close up view of the resistance, normalized to \textit{R}(375 K), around \textit{T}$_3$ for \textit{x} = 0.0026. The \textit{R}(\textit{T}) is hysteretic between cooling and warming data implying \textit{T}$_3$ is first-order. (b) Residual resistance ratios for all Mn(Pt$_{1-x}$Pd$_x$)$_5$P samples (note the log scale for the y-axis).}
    \label{RT_hysteresis_RRR}
\end{figure}

Table \ref{EDS_SC} lists the nominal compositions Mn$_9$Pt$_{71-y}$Pd$_y$P$_{20}$ used for the growth of Mn(Pt$_{1-x}$Pd$_x$)$_5$P single crystals and the corresponding values of \textit{x} determined by EDS from each batch. Note that starting compositions do not correspond to exact Mn(Pt$_{1-x}$Pd$_x$)$_5$P stoichiometry (i.e. \textit{y} $\neq$ \textit{x}) because the intention for solution growth is to intersect the liquidus surface for crystalization of Mn(Pt$_{1-x}$Pd$_x$)$_5$P on cooling, not to be directly "on-line". Table \ref{EDS_SC} gives the same information for the polycrystalline samples obtained from solid-state reactions. The values of \textit{x} are the average of multiple scans obtained on each sample and the error bars were determined either by the standard deviations or the EDS fitting error (the fitting error was only used for \textit{x} = 0.0001). We calculated \textit{x} relative to the total amount of Pt and Pd detected in each sample, i.e. \textit{x} = \textit{f}$_{\text{Pd}}$/(\textit{f}$_{\text{Pt}}$+\textit{f}$_{\text{Pd}}$) where \textit{f}$_{\textit{Pt}}$(\textit{f}$_{\textit{Pd}}$) represent the quantity of Pt(Pd) found in each sample. For both the solution growth single crystals and polycrystalline samples, we observe a monotonic enrichment of the EDS Pd fraction as the starting growth compositions become richer in Pd. For the single crystalline sample with nominal \textit{y} = 0.25 and EDS \textit{x} = 0.00013, the Pd fraction is under the detection limit of our instrument (the fitting errors to the EDS spectra were $\approx$ 10 times greater than the detected quantity of Pd). This is unsurprising given the extremely dilute quantity added to the initially melt (\textit{y} = 0.25, which is $\approx$ 0.35 $\%$ Pd relative to the total quantity of Pt and Pd); however, the significant differences in the magnetic and transport properties between this sample and those of pure MnPt$_5$P imply very small but finite Pd incorporation (see Figures \ref{MT}--\ref{MH_lowx} in the main text). Given the detected Pd in this sample is below the resolution of our instrument, to estimate the Pd fraction, we extrapolated the linear trend between nominal and EDS values of \textit{x} to nominally \textit{x} = 0.0035 (see the inset to Figure \ref{xnomxEDS}), which gives an estimate of \textit{x} = 0.0013 for the most dilute sample.

\subsection{Crystallographic information for Mn(Pt$_{1-x}$Pd$_x$)$_5$P}

Table \ref{Mn(Pt1-xPdx)5P_crystallography} gives the refinement information and statistics for the single crystal XRD refinements of Mn(Pt$_{1-x}$Pd$_x$)$_5$P samples. The single crystals used for these measurements were picked from the sintered pellets (solid-state reactions). The atomic positions and isotropic displacement parameters are listed in Table \ref{Mn(Pt1-xPdx)P_refineatoms}. The results indicate that MnPd$_5$P and the Mn(Pt$_{1-x}$Pd$_x$)$_5$P compounds all adopt the layered tetragonal (\textit{P}4/\textit{mmm}) anti-CeCoIn$_5$ type structure. The single crystal XRD refinements support mixed occupancy between Pt and Pd on the two atomic sites 1\textit{a} and 4\textit{i}, indicating the formation of a Mn(Pt$_{1-x}$Pd$_x$)$_5$P solid solution. As shown in Table \ref{Mn(Pt1-xPdx)P_refineatoms}, the Pd atoms may have a slight preference for occupying the 4\textit{i} site over the 1\textit{a} site; however, given the uncertainties of our experiments we are not able to confidently assert at sight preference for the Pt or Pd atoms.

\subsection{Additional Data for Mn(Pt$_{1-x}$Pd$_x$)$_5$P Single Crystals}

Figure \ref{MH_highx} shows the magnetization isotherms measured at 2 K for Mn(Pt$_{1-x}$Pd$_x$)$_5$P where 0.022 $\geq$ \textit{x} $\geq$ 0.01. All datasets show field dependencies that are characteristic of ferromagnetic order in which the \textit{ab}-plane is the easy direction (i.e. \textit{H} $\perp$ \textit{c}). The saturation magnetization are all on the order of 4-4.5 $\mu_{\text{B}}$/f.u., with $\approx$ 10 $\%$ variation amongst the samples, and there is no clear trend in the magnitude of $\mu_{\text{sat}}$ as a function of the Pd fraction. The discrepancies likely represent a combination of weighing errors, uncertainty in \textit{x}, and intrinsic changes in $\mu_{\text{sat}}$ as \textit{x} changes. Ultimately, the primary conclusion drawn from Figure \ref{MH_highx} is that the \textit{x} $>$ 0.01 Mn(Pt$_{1-x}$Pd$_x$)$_5$P are unambiguously easy-plane ferromagnets below \textit{T}$_1$.

\begin{figure}[!t]
    \centering
    \includegraphics[width=\linewidth]{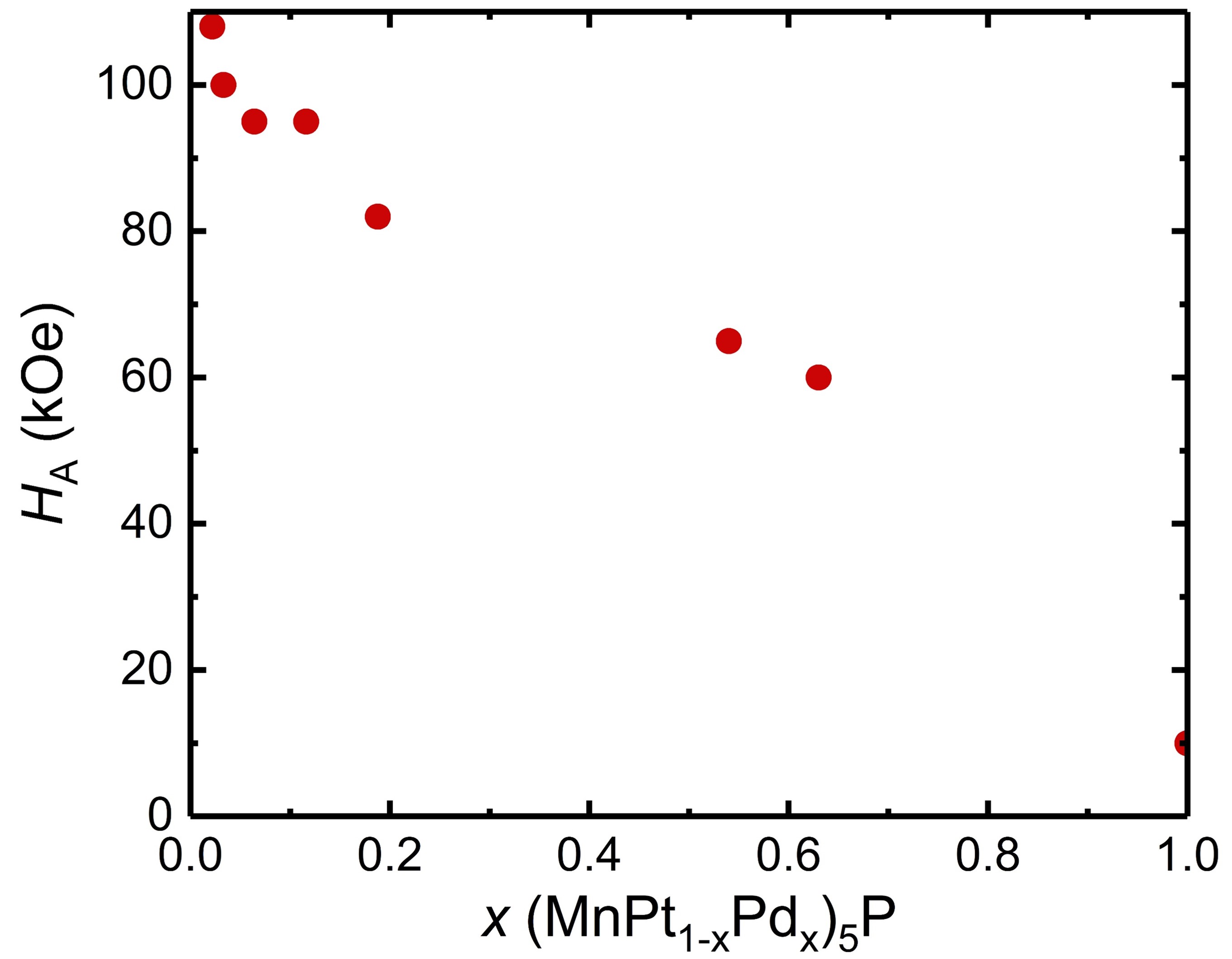}
    \caption[]{Anisotropy fields for Mn(Pt$_{1-x}$Pd$_x$)$_5$P single crystals estimated from the extrapolations of the tangents in Figure \ref{MH_highx}.}
    \label{HA}
\end{figure}

Extrapolating the tangents of the \textit{M}(\textit{H}) isotherms in Figure \ref{MH_highx} to their intersection point gives an estimate of the anisotropy fields \textit{H}$_{\text{A}}$, which are plotted explicitly in Figure \ref{HA}. We find that \textit{H}$_{\text{A}}$ decreases monotonically with increasing Pd fraction, strongly suggesting that the weakening mangetic anisotropy is governed by the decreasing strength of spin orbit coupling accompanying the replacement of Pt atoms with lower \textit{Z} Pd. 

Figure \ref{RT_hysteresis_RRR}a shows the temperature dependent resistance for \textit{x} = 0.0026 zoomed in around \textit{T}$_3$, i.e., the lower temperature re-entry into the ferromagnetic state. The resistance is hysteretic between cooling and warming temperature sweeps, showing that \textit{T}$_3$ is first-order. 

Figure \ref{RT_hysteresis_RRR}b displays the residual resistance ratios (\textit{RRR} = \textit{R}(375 K)/\textit{R}(1.8 K)) of the Mn(Pt$_{1-x}$Pd$_{1-x}$)$_5$P samples. We find that the \textit{RRR} is highest for \textit{x} = 0 and \textit{x} = 1 and decreases rapidly when the composition moves away from the pure phases, reaching a minimum value of $\approx$ 3.5 at \textit{x} = 0.54. The reasonably high \textit{RRR} values for the pure compounds (115 for \textit{x} = 1 and 37 for \textit{x} = 1) indicates high crystal quality, and the swift reduction of \textit{RRR} as \textit{x} approaches 0.5 is consistent with stronger scattering of charge carriers accompanying the increasing crystallographic disorder between Pt and Pd atoms in the alloys.

\subsection{Magnetic Data for Polycrystalline Mn(Pt$_{1-x}$Pd$_x$)$_5$P}

\begin{figure}[!t]
    \centering
    \includegraphics[width=\linewidth]{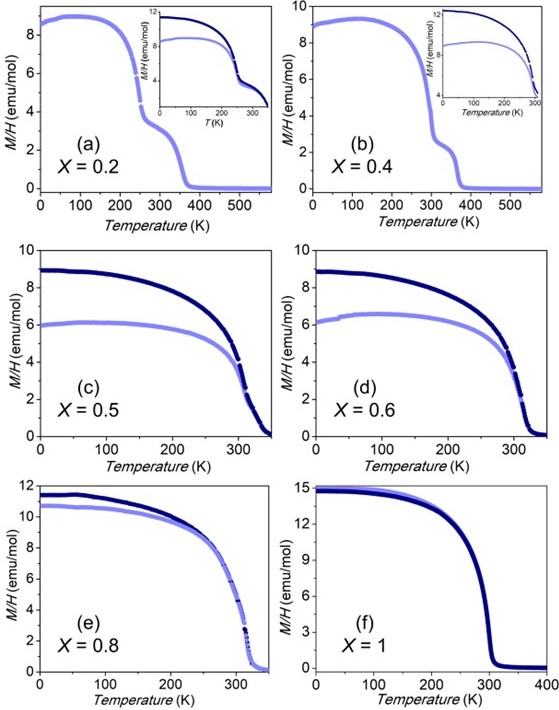}
    \caption[]{Temperature dependent \textit{M/H} measured on pellets of polycrystalline Mn(Pt$_{1-x}$Pd$_x$)$_5$P.}
    \label{MT_pc}
\end{figure}

\begin{figure*}[!t]
    \centering
    \includegraphics[width=\linewidth]{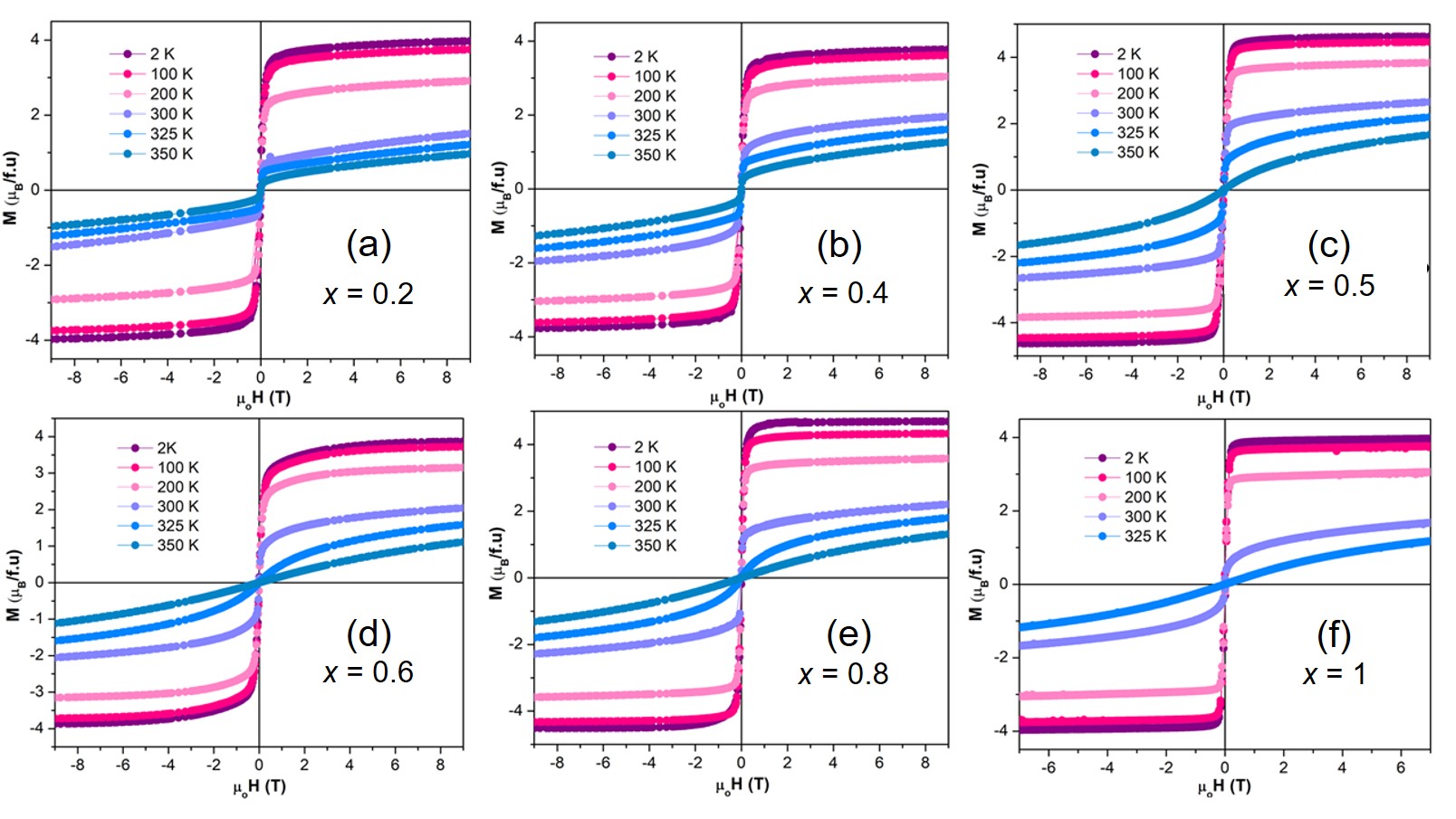}
    \caption[]{Field dependent magnetization isotherms measured at 2 K on pellets of polycrystalline Mn(Pt$_{1-x}$Pd$_x$)$_5$P.}
    \label{MH_pc}
\end{figure*}

Because the crystallographic data for MnPd$_5$P and Mn(Pt$_{1-x}$Pd$_x$)$_5$P was obtained from small single crystals picked from solid state reactions, we also measured the magnetic properties of the sintered pellets to ensure consistency with the solution-grown single crystals discussed in the main text. The magnetic properties of the polycrystalline Mn(Pt$_{1-x}$Pd$_x$)$_5$P (\textit{x} = 0.2, 0.4, 0.5, 0.6, 0.8, and 1) were measured in a Quantum Design PPMS Dynacool (QD-PPMS) at the National High Magnetic Field Laboratory over a temperature range of 1.8 to 400 K with the applied field of 1 kOe. Additionally, magnetic measurements of Pt doped compounds were carried out in a vibrating sample magnetometer (VSM) in a Quantum Design PPMS system over a temperature range of 1.8–600 K with the applied field of 1 kOe. The field dependent magnetization measurements were carried out at several different temperatures between 2–350 K and in fields up to 90 kOe.  

The Magnetic data  for the polycrystalline Mn(Pt$_{1-x}$Pd$_x$)$_5$P samples is displayed in Figures \ref{MT_pc} and \ref{MH_pc}. Like the single crystals discussed above, the Pd-containing polycrystalline samples, which have 0.2 $\geq$ \textit{x} $\geq$ 1, also show ferromagnetic behavior where \textit{T}$_C$ is maximized near 312 K for $\approx$ \textit{x} = 0.60 Pd. However, in addition to the Mn(Pt$_{1-x}$Pd$_x$)$_5$P primary phase, the polycrystalline pellets also contained a small fraction of a ferromagnetic impurity, likely MnPt$_3$ (\textit{T}$_C$ $\approx$ 390 K).\autocite{PhysRev.187.611,ANTONINI1969310} Owing to the strong response of ferromagnetism to an applied magnetic field, even small ferromagnetic impurities are easily detected in magnetization measurements, and our polycrystalline samples with \textit{x} $<$ 0.5 all show high temperature (\textit{T} $>$ 300 K) ferromagnetic-like transitions that are not observed in any of the datasets collected on the single crystals. An analogous MnPd$_3$ phase also exists, but orders antiferromagnetically,\autocite{PhysRev.139.A1581,KREN1969340,KREN19721195} which likely explains why the polycrystalline Mn(Pt$_{1-x}$Pd$_x$)$_5$P samples with \textit{x} $>$ 0.5 do not show evidence for a second transition. As the ordering of MnPt$_3$ could easily be misinterpreted as a second, higher temperature, transition in the Pt-rich samples, the contrast between single and polycrystalline data is an excellent demonstration of the advantages of solution growth, which allows us to produce high quality Mn(Pt$_{1-x}$Pd$_x$)$_5$P crystals free of significant contamination by magnetic impurities. The field dependent magnetization isotherms for the polycrystalline samples are shown in Figure \ref{MH_pc} and show soft ferromagnetic behavior below \textit{T}$_1$ and saturated moments $\approx$ 4-4.5 $\mu_{\text{B}}$/f.u. at 2 K.  Excluding the transitions from the MnPt$_3$, the intrinsic Curie temperatures of the polycrystalline samples are otherwise in good agreement with those inferred from the data collected on single crystals.

\printbibliography

\end{singlespace}

\end{document}